\documentclass[a4paper,10pt]{article}
\usepackage[a4paper, lmargin=1.0in, rmargin=1.0in, tmargin=1.0in, bmargin=1.0in]{geometry}

\usepackage{graphicx}
\usepackage{amsmath,amssymb}

\def \imply {\Rightarrow}
\def \ep {{\epsilon}}

\newcommand{\define}{\widehat{=}}
\newcommand{\ra}{{\rightarrow}}
\newcommand{\lan}{{\langle}}
\newcommand{\ran}{{\rangle}}
\newcommand{\cM}{{\cal M}}

\newcommand{\bigc}{\bigcirc}

\newcommand{\cB}{{\cal B}}
\newcommand{\T}{{\cal T}}
\newcommand{\X}{{\cal X}}
\newcommand{\N}{{\mathbb N}}
\newcommand{\C}{{\cal C}}
\newcommand{\U}{{\cal U}}
\newcommand{\V}{{\cal V}}

\newcommand{\mbR}{{\mathbb R}}
\newcommand{\mbN}{{\mathbb N}}

\newcommand{\comp}[1]{\ensuremath{[\!\!| #1 |\!\!]}}
\newcommand{\dig}[1]{\ensuremath{[ #1 ]}}

\newcommand{\Tr}{{\rm Tr}}
\newcommand{\true}{{\tt true}}
\newcommand{\false}{{\tt false}}

\newtheorem{Theorem}{Theorem}[section]
\newtheorem{Fact}[Theorem]{Fact}

\title{A Simplification of a Real-Time Verification Problem}
\author{Indranil Saha$^1$ \and Janardan Misra$^2$ \and Suman Roy$^3$\\
     $^1$ Comp. Sc. Dept., Uni. of California, Los Angeles, CA 90095, USA.\\ Email:indranil@cs.ucla.edu\\
     $^2$ HTS Research, BG Road, Bangalore 560076, India.\\ Email: janmishra@gmail.com \\
     $^3$ SETLABS, Infosys Tech. Ltd., \#44 Electronic City, Bangalore 560100, India.\\ Email: suman\_roy@infosys.com}

\begin{document}
\maketitle

\begin{abstract}
We revisit the problem of real-time verification with dense time dynamics using timeout and calendar based models, originally proposed by Dutertre and Sorea, and simplify this to a finite state verification problem. To overcome the complexity of verification of real-time systems with dense time dynamics, Dutertre and Sorea, proposed timeout and calender based transition systems to model the behavior of real-time systems and verified {\tt safety} properties using $k$-induction in association with bounded model checking. In this work, we introduce a specification formalism for these models in terms of Timed Transition Diagrams and capture their behavior in terms of semantics of Timed Transition Systems.  Further, we discuss a technique, which reduces the problem of verification of qualitative temporal properties on infinite state space of (a large fragment of) these timeout and calender based transition systems into that on clockless finite state models through a two-step process comprising of digitization and canonical finitary reduction. This technique enables us to verify {\tt safety} invariants for real-time systems using finite state model-checking avoiding the complexity of infinite state (bounded) model checking and scale up models without applying techniques from induction based proof methodology. Moreover, we can verify {\tt liveness} properties for real-time systems, which is not possible by using induction with infinite state model checkers. We present examples of Fischer's Protocol, Train-Gate Controller, and TTA start-up algorithm to illustrate how such an approach can be efficiently used for verifying {\tt safety}, {\tt liveness}, and {\tt timeliness} properties specified in LTL using finite state model checkers like {\sf SAL-smc} and {\sf Spin}. We also demonstrate how advanced modeling concepts like inter-process scheduling, priorities, interrupts, urgent and committed location can be specified as  extensions of the proposed specification formalism, that can be subjected to the proposed two step reduction technique for verification purposes.
\end{abstract}


\tableofcontents

\section{Introduction}
Real-time systems are an important class of mission critical systems, which have been well studied for their design, implementation, performance and verification. 
Modeling and verification of real-time systems in dense time domain is an important problem area that evoked lot of research interest in the recent past. Because of the fact that the state space of real-time systems with continuous dynamics is uncountable, modeling and verification of them is rather difficult, in particular using explicit state model checkers. Many formalisms have been used to model and verify real-time systems. Notable among them are different kinds of timed transition models~\cite{Alur99, HMP92a},  timed process algebras~\cite{BB91,DS95,NS94}, and real-time logics~\cite{AH91,BMN00}.

In~\cite{DS04a}, Dutertre and Sorea, considered verification of a train-gate controller modeled as a timed automata. Though they could specify the timed automata model in terms of state transition system in infinite state model checker {\sf SAL}~\cite{MOR+04}, it however did not to produce the desired results. In particular, the clock variables occurring in timed automata would be updated in arbitrarily small increments leading to infinite trajectories during which the discrete state remained idle. This made proof of {\tt safety} properties by $k$-induction quite hard, and sometimes impossible. The fact that the traditional semantics of timed automata allows several time steps to occur in succession is an obstacle in proving properties by $k$-induction.

To address this problem the same authors proposed timeout and calender based transition models,~\cite{DS04a,DS04b}, originally from discrete event simulation, to represent the behavior of timed triggered systems with dense time dynamics. These models are amenable to general-purpose verification environments, like {\sf SAL} in which state machines and their compositions can be specified. In this modeling approach, each process in the system has a timeout that holds the time when the next discrete transition of the process would happen, and there is a global data structure, called \textit{calendar}, which stores future events (message delivery) and the time points at which these events are scheduled to occur. During the time progress transition, time is advanced to the minimum of timeouts of processes, or to the least time point at which a message will be delivered in future, whichever is less. Further, Dutertre and Sorea, used this calendar based model along with timeouts for individual processes to model TTA startup protocol in {\sf SAL}~\cite{DS04b}. Using bounded model checking, they proved a {\tt safety} property by $k$ induction. However, these proofs using $k$-induction do not usually scale up well; a {\tt safety} property often cannot be proved at induction depth $1$. Sometimes {\tt safety} properties are not at all inductive and need the support of auxiliary lemmas. In~\cite{DS04b}, a {\tt safety} property for the TTA startup algorithm with only $2$ nodes has been proved by using $3$ additional lemmas. A verification diagram based abstraction method proposed in~\cite{Rus00} has been used to prove the same {\tt safety} (invariant) property for scaled up models (upto $10$ nodes). However, {\tt liveness} properties still remain beyond the scope of this approach.

While only {\tt safety} properties can be verified on these models with dense time, discrete time modeling of the same can help verify {\tt liveness} and {\tt timeliness} properties, and also help scale up proofs for {\tt safety} properties.
It turns out that verification of a real-time system in dense domain is equivalent to verifying the system in discrete domain if both the behavior of the system captured by the model and the properties considered are digitizable~\cite{HMP92}. It can be shown that if the timeout updates are not restricted to $(0,1)$-intervals, then similar to the timed transition system of~\cite{HMP92} (refer to theorem 2 therein), transition systems for timeout and calendar based models also give rise to digitizable behaviors (computations). Also verification of qualitative properties like {\tt safety} and {\tt liveness}, in discrete time domain is equivalent to verifying these properties in dense time domain (refer to proposition $1$ in~\cite{HMP92}).

Techniques like bounded model checking~\cite{MRS03,DS04a} can be useful for detecting bugs during the verification process even in discrete domain, where  one systematically searches for counterexamples of length bounded by some integer $k$. The bound $k$ is increased until a bug is found, or some pre-computed completeness threshold is reached. Unfortunately, it is usually very expensive to compute completeness thresholds. Also these thresholds may be too large to effectively explore the bounded search space. Additionally, such completeness thresholds may be absent for many infinite-state systems. A finite state modeling of the system can help exploring the state space much easily. Examples of finite state model checkers are {\sf Spin}~\cite{Hol03}, {\sf SAL-smc} solvers~\cite{DS04a} etc. {\sf Spin} has been used to finitely model TTA startup algorithm using a clockless calendar based model~\cite{SMR07}.  In terms of  scalability, finite state verification of TTA in {\sf Spin} is almost comparable to the verification of TTA based on verification diagram oriented abstraction method~\cite{DS04b}. Moreover, {\tt liveness} properties can be verified in this framework.

\begin{figure}
\centering
\includegraphics[scale=0.7]{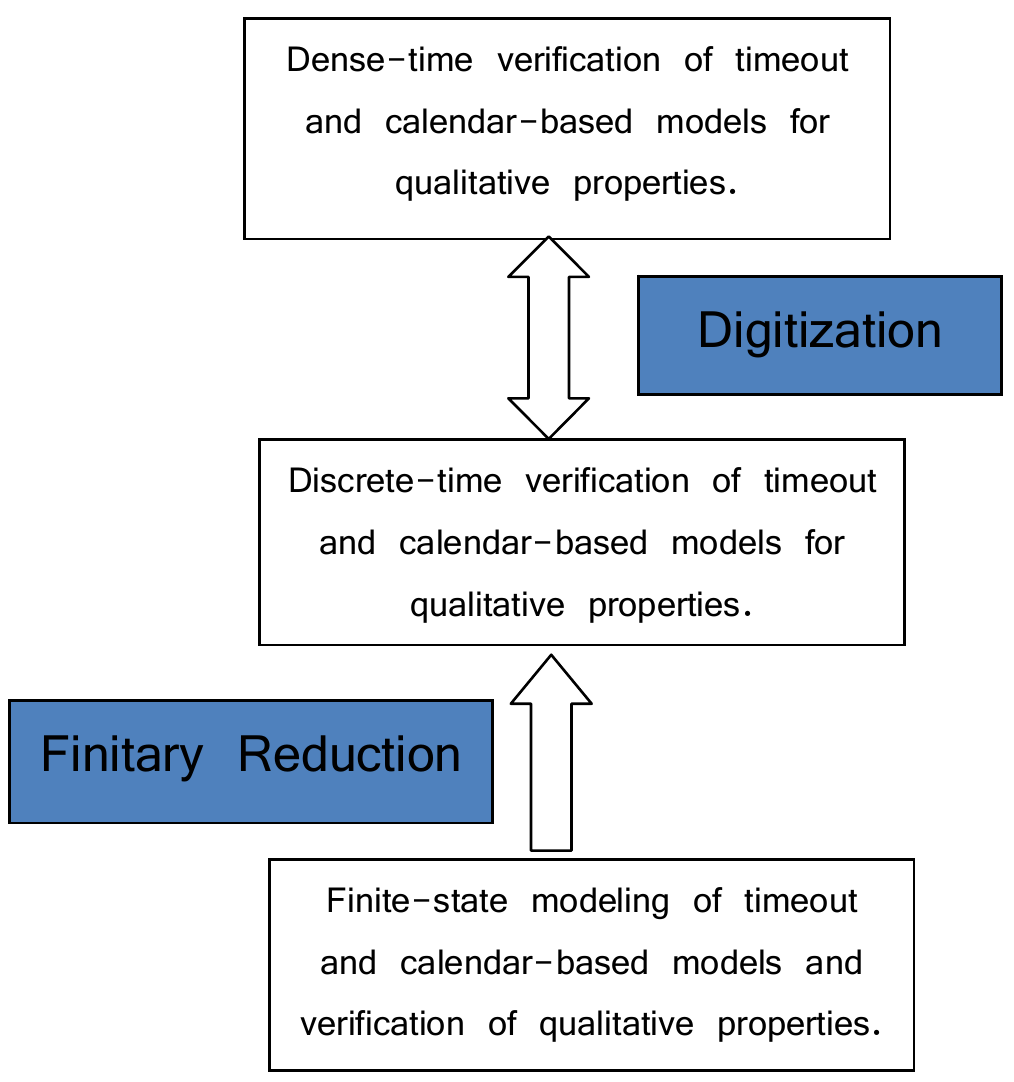}
\caption{A two-step verification process } \label{fig:fig0}
\end{figure}

In this work, we aim to carry out a finite state modeling and verification on timeout and calendar models without continuously varying clocks. As there are drawbacks of those models earlier proposed from the point of view of design considerations, like absence of formally defined syntactic models and associated semantics, we slightly deviate from them.
We consider the specification framework of timed transition diagrams and extend it to formalize timeout and calendar based models as timeout and calendar based transition diagrams and their behavior in terms of semantics of transition systems.
The benefits that we derive from using this formalization are  many-fold. Our framework of timeout transition diagrams inherits most of the properties of classical timed transition system introduced in~\cite{HMP92a}. Most of the techniques, like digitization that can be applied to these timed transition systems are applicable to our formalization also. This can be also used to model time-triggered systems and reason about them. Finally we use this formal modeling framework to reduce continuous time verification problem to discrete time finite state verification, albeit under some restrictions. Towards that, we use a two step technique comprising of digitization and finitary reduction (a schematic diagram of this technique is shown in Figure~\ref{fig:fig0}). We show that the computations of timeout and calendar models are digitizable provided the timeout increments are not restricted to $(0,1)$-interval.
As LTL properties are qualitative and hence, are digitizable, verification of LTL properties on timeout and calendar models in dense time is equivalent to that in discrete time.
The next step is to reduce this problem into an equivalent finite state verification problem. We could not directly proceed to extract finite state models from dense time models, since the latter models are inherently infinite (and dense) and hence it is not possible to render them finite even by bounding the variables. Also note that such a modeling cannot be directly subjected to finite state verification since for timeout and calendar based models, global time and timeouts always increase. Nonetheless, we propose a finitary reduction technique which effectively reduces the infinite state timeout and calendar based transition systems with discrete dynamics into a finite state transition system. We achieve this by using a clockless modeling technique which effectively strips the model of the global clock and keeps track of the relative updation of timeouts, and restricts the values of variables/timeout updates to bounded domains. We demonstrate by examples, how such a modeling approach can be efficiently used for verifying {\sf safety}, {\sf liveness}, and
{\sf timeliness} properties using 
finite state model checkers, {\sf SAL-smc} and {\sf Spin}. We also highlight the scalability of such models for verification purposes by comparing the performance of such models under dense time and finite state modeling. A preliminary version of this paper appeared in~\cite{SMR07}.

The remainder of the paper is organized as follows. In the next section, we briefly discuss the timeout and calendar based modeling as presented in~\cite{DS04a,DS04b}. In Section~\ref{formalize}, we
present the formalization of these models in terms of timeout transition diagrams and their behavior in terms of the semantics of transition systems. We discuss the technique of digitization in Section~\ref{digitize} and present our first step of reduction of dense-time verification problem to integral time verification problem. In Section~\ref{clockless}, we describe the finitary reduction technique and subsequently, formalize it in terms of clockless modeling. 
We present experimental results in Section~\ref{results}. A few extensions of our framework are described in Section~\ref{extend} followed by concluding remarks in Section~\ref{conclude}.

\subsection{Related Work}
There have been earlier attempts to model and verify time-triggered systems using extensions of finite state model checkers, e.g., {\sf Spin}. {\sf Spin}~\cite{Hol03} is a tool for automatically model checking distributed systems, but it does not allow explicit representation for time. There are mainly two attempts for extending {\sf Spin} with time. Real-time extension of {\sf Spin} ({\sf RT-Spin}~\cite{TC96}) is one such attempt, that makes use of timed Buchi automata~\cite{AD94} with real-valued clocks as a modeling framework.
However, this formalism is incompatible with the partial order reduction which is supported by {\sf Spin}. Another is the work on {\sf DT-Spin}~\cite{BD98a,BD98b}, which allows one to quantify (discrete) time elapsed between events, by specifying the time slice in which they occur. {\sf DT-Spin} is compatible with the partial order reduction and has been used to verify industrial protocols, like, AFDX Frame management protocol~\cite{SR06a} and TTCAN~\cite{SR06b}. Nonetheless, systems with asynchronous communication with bounded delays between components cannot be modeled directly by using the kind of asynchronous channels that {\sf Spin} provides, since there is no explicit provision to capture message transmission delays. A possible way is to model each channel as a separate process with delay as a state variable. In~\cite{BD98a}, the channels in the
example of PAR protocol have been implemented in the same way. But
for systems with relatively large number of components and high
degree of connectivity among them, modeling channels in this way is
difficult, and hence state space explosion becomes an unavoidable
problem. 

The concept of clockless modeling has been introduced in~\cite{pike05}. In that Pike
builds on the work of~\cite{DS04a}  and proposes a new formalism called Synchronizing Timeout Automata (STA) to
reduce the induction depth $k$ required for $k$-induction. He
introduces a clockless semantics for STA so that the resulting
transition system does not involve a clock. STA in effect, describes
the overall system architecture in terms of timeout transition
system introduced in~\cite{DS04a}. A closer analysis of the {\sf SAL}
model for the example of Train-Gate Controller presented
in~\cite{pike05}, reveals that the considered model is not
deadlock free. This is because the model fails
to specify the timeout updation rules precisely for the transitions leading to
a waiting state. When a process is waiting for an external
signal, its timeout should be set to a value greater than the
current value of the timeouts of the senders of the expected signal.
This kind of modeling errors could possibly be eliminated with a suitable modeling framework such as the one proposed in this paper.

To our knowledge, the first attempt to convert TA to untimed TA is taken up in~\cite{AD94}. Building upon these, in~\cite{CH04} the authors discuss how a special kind of model for specifications written in Duration Calculus (DC)~\cite{dc91} can be generated in which, DC formulas would correspond to regular expressions over a state of special symbols. The models for DC formulas contain discrete states and digitization of continuous states,
thereby enabling reasoning in a single framework of both discrete and continuous time. Applying discretization on the continuous component of real-time systems, these models could be further translated into Promela models for verification experiments using SPIN.

\section{Timeout and Calendar-based Real-Time Models}\label{basic-framework}

In this section we briefly discuss the timed automata~\cite{AD94}, timeout, and calendar-based models introduced earlier in~\cite{DS04b}.

\subsection{Timed Automata}

Timed automata (TA) was introduced by Alur et al. in~\cite{AD94} as a clock based model for specifying real-time system designs. TA is widely used for modeling and verification of real-time systems. Many tools are available for analyzing timed automata e.g., UPPAL~\cite{BDL04}, Kronos~\cite{BDM+98}, Rabbit~\cite{BLN03}. For further details on TA, the reader is referred to~\cite{AD94}.

\subsection{Timeout Transition Model}

Dutertre and Sorea~\cite{DS04a,DS04b} used timeout based modeling to
formally verify real-time systems using $k-$induction in {\sf SAL} model checker.
A Timeout Transition Model (TTM), which is a model of the combined system behavior, contains a finite set of timeouts and a global clock variable $t$.
Timeouts define the time points when discrete transitions will be enabled
in the future. The clock variable $t$ keeps track of the current
time. In practice a typical real-time system may contain a number
of processes. Every process is associated with one timeout which
records the future point of time when the next discrete transition
for the process is scheduled to occur. Transitions in this model
are classified into two types: time progress transitions and
discrete transitions. In time progress transition, $t$ (time)
advances to the minimum valued timeout(s). Discrete transition
occurs when $t$ is equal to the minimum valued timeout(s). If
there are more than one processes, which have their timeouts equal
to the minimum value, one of them is randomly chosen and
corresponding discrete transition occurs updating the value of the
timeout for the selected process. Timeout based modeling approach
is suited to model systems where the processes communicate via
shared variables or the communication between the processes is a
rendezvous one.

\subsection{Calendar Transition Model}

Interprocess communication delay during message transfers cannot be modeled using
timeout based modeling because delays are beyond the control of
individual processes. Addition of an {\it event calendar}, a globally shared
data structure, is proposed as a convenient way to model
such delays~\cite{DS04b}. This model is called Calendar Transition Model (CTM). A {\it calendar} is a set of bounded size of the form $C = \{
\lan e_1, t_1 \ran, \ldots, \lan e_r, t_r \ran \}$, where each event $e_i$
is associated with the time point $t_i$ when it is scheduled to
occur. There is fundamental difference between a clock and a calendar in the sense that while the former measures the time elapse since its last reset, the latter stores expected delivery delays for all undelivered messages. Asynchronous communication with bounded delay can be easily modeled by using calendar as a global data structure. When a message is transmitted by a process, it is added to the calendar as an event $e_i$ to occur at time $t_i$, where $t_i$ denotes the expected delivery time for the message. On receiving the message, the event is removed from the calendar. Thus at any state, the calendar $C$ can be seen as a set of messages that have been sent but are yet to be received with corresponding expected delivery delays.

\subsection{Limitation of Existing Formalisms}

Timed automata is one of the most frequently used formalism for specifying real-time system designs. However as it turns out that for systems with asynchronous communication with bounded delays between components TA does not offer any efficient means of specification. Two possible choices have been considered in literature. First choice is to use state variables for encoding the behavior of asynchronous channels however without any explicit provision to capture message transmission delays. Second choice is to model each channel as a separate TA with delay as a state variable. However with relatively large number of components and high degree of connectivity among them, modeling channels in this way is difficult, and state space explosion becomes an unavoidable problem. UPPAAL~\cite{BDL04}, which can model TA, has the same problem when it is used to model asynchronous communications with bounded delays - every channel has to be modeled as a separate TA capturing the message transmission delays. 

On the other hand, although TTM and CTM are expressive enough to capture a range of behaviors associated with time triggered systems including asynchronous communication delays, they however have two specific design limitations:
\begin{itemize}
\item These models are not well suited for actual system design purpose since they describe the behavior of the combined system without (explicitly) specifying the design of the modular components.
\item Absence of formally defined syntactic design models corresponding to these transitions systems would demand that additional correctness measures are put in place because for verification purposes actual designs models need to be (manually) interpreted and translated into these transition systems as per the underlying system dynamics and on discovering an error during verification, such errors need to comprehended by a designer, and subsequently, translated back into his design for a remedial action.
\end{itemize}

Keeping in view of such limitations in the existing specification formalisms, we will next define and elaborate using examples a new timeout based formalism, which can effectively overcome these barriers. 




\section{Formalization of Timeout and Calendar based Models}\label{formalize}

In~\cite{HMP92a} an abstract model of timed transition system was proposed which could
represent a wide variety of behaviors of the timed execution of concurrent processes. In this section we adapt and extend the Timed Transition System (TTS) described therein to represent timeout and calendar based models. Further we describe their associated semantics in terms of state transition systems.

\subsection{Timeout based Timed Transition Model}

\subsubsection{Syntax}\label{tom_syntax}

A Timeout based Model (ToM) is represented as $$P:\ \{\theta\}[P_1 || P_2 ||\ldots ||P_n].$$ Each process $P_i$ is a sequential non-deterministic process having $\tau_i$ as its local timeout and $\X_i$ as a set of local timing variables. Local timing variables are used for determining the relative delay between events. A shared variable $\{t\}$ represents the global clock. The operator ``$||$'' denotes parallel composition. The formula $\theta$, called the {\em data pre-condition} of $P$, restricts the initial values of variables in $$\U = \{t\} \cup \T \cup \X \cup \mathit{Var},$$ where the set of all timeouts is $\T = \{\tau_1, \tau_2, \ldots, \tau_n\}$, and $\X = \bigcup_i \X_i$. The set $\mathit{Var} = (G\ \cup\ L_{1}\cup L_{2}\cup\ldots\cup L_{n})$ is the set of other state variables. The variables in $G$ are globally shared among all the processes while $L_{i}$ contains  variables local to process $P_i$. Let $f^{\mathit{Var}}$ be the set of computable functions on $\mathit{Var}$.

Each process $P_i$ is represented using a \emph{timeout transition diagram} (TTD), which is a finite directed graph with a set of nodes $\mathit{Loc}_i = \{l_0^i, l_1^i,\ldots,l_{m_i}^i\}$, called \emph{locations}. The entry location is $l_0^i$. There are two kinds of edges in the graph of a process $P_i$: {\it Timeout edges} and {\it Synchronous Communication edges}. Edge definitions involve an enabling condition or guard $\rho$, which is a boolean-valued function or a predicate.\\\\
{\sf Timeout Edges:} A timeout edge $(l_j^i, \rho \Rightarrow \lan \tau_i := update_i, \gamma, f \ran, l_k^i)$ in the graph of the process $P_i$ is represented as $$l_j^i\ \stackrel{\rho\ \Rightarrow \lan \tau_i := update_i, \gamma, f \ran}{\longrightarrow}\ l_k^i,$$ where $update_i$ specifies the way timeout $\tau_i$ is to be updated on taking a transition on the edge when the guard $\rho$ evaluates to $\true$.  $\gamma \subseteq \X_i$ specifies the local timing variables which capture value of the clock $t$ while taking transition on the edge. This value may be used during future transitions while estimating relative delay w.r.t. this transition. $f \in f^{\mathit{Var}}$ manipulates the state variables in $G \cup L_i$.

$update_i$ is defined using the rule: $$update_i = k_1\ |\ k_2\ |\ \infty\ |\ \max(\cM)$$ where $l + z \prec k_1 \prec' m + z'$ for $\prec,\prec' \in \{<, \leq\}$ and $k_2 \succ l + z$ for $\succ \in \{>, \geq\}$; $z, z' := t | w$ and $l, m \in \N$ are non negative integer constants specifying the lower and upper limits for a timeout increment interval\footnote{This interval mimics the delay interval marking an edge in the original timed transition diagrams}, and $w \in \X_i$ is a local timing variable. The variable $z$ makes such an interval relative to the occurrence of specific events. ${\cM}$ is the set of all the integer constants that are used to define the upper limit of different timeouts for different processes in the system. $\max(\cM)$ returns the maximum of all the integers in $\cM$.


Constraints on $k_1, k_2$ specify how new value of timeout $\tau_i$ should be determined based upon the current value of the clock $t$ and/or $w$, which would have captured the value of $t$ in some earlier transition.
Setting a timeout to $\infty$ tends to capture the requirement of indefinite waiting for an external signal/event.
The selection of the timeout value using $\max(\cM)$ is used to capture the situation where the next discrete transition of a process may happen at any time in the future, for example, the process may be in a sleeping mode and can wake up at any future point of time.\\ \\
{\sf Synchronous Communication Edges:} Rendezvous communication between a pair of processes $(P_s, P_r)$ is represented by having an edge pair $(e_s, e_r)$ s.t. $e_s \in P_s$ and $e_r \in P_r$ and $$e_s: l_j^s\ \stackrel{\rho\  \Rightarrow \lan ch!m, \tau_s := update_s, \gamma, g \ran}{\longrightarrow}\ l_k^s$$ $$e_r: l_j^r\ \stackrel{True\  \Rightarrow \lan ch?\bar{m}, \tau_i := update_r, \gamma', h \ran}{\longrightarrow}\ l_k^r$$ where $ch$ is the channel name, $m \in L_i$ is the message sent, and $\bar{m} \in L_r$ the message received, and $g, h \in f^{\mathit{Var}}$ are the computable functions.

\subsubsection{Semantics}~\label{tts}

With a given ToM $$P:\ \{\theta\}[P_1 || P_2 ||\ldots ||P_n]$$ we associate the following transition system $S_P = (\V, \Sigma, \Sigma_0, \Gamma)$, referred to as {\it timeout based clocked transition system} (TCTS) where,

\begin{enumerate}
    \item $\V = \U \cup \{\pi_1, \ldots,\pi_n\}$. Each \emph{control variable} $\pi_i$ ranges over the set
    $\mathit{Loc}_i \cup \{\perp\}$. The value of $\pi_i$ indicates the location of the control for the process $P_i$ and it is $\perp$ (undefined)
     before the start of the process.
    \item $\Sigma$ is the set of states. Every state $\sigma \in \Sigma$ is an interpretation of $\V$, that is, it assigns values to clock variable $t$, every timeout variable in $\T$, timing variables in $\X$, state variables in $\mathit{Var}$, and control variables $\pi_1, \ldots, \pi_n$, in their respective domains. For $x \in \V$, let $\sigma(x)$ denote its value in state $\sigma$.
\item $\Sigma_0 \subseteq \Sigma$ is the set of initial states such that for every $\sigma_0 \in \Sigma_0$, $\theta$ is true
in $\sigma_0$ and $\sigma_0(\pi_i) = \perp$ for each process
$P_i$.
\item $\Gamma = \Gamma_e \cup \Gamma_+ \cup \Gamma_0 \cup \Gamma_{syn\_comm}$ is the set of transitions. Every transition
$\nu \in \Gamma$ is a binary relation on $\Sigma$ defined further as follows:
\end{enumerate}
{\sf Entry Transitions: } $\Gamma_e$, the set of entry
transitions contains an \emph{entry transition} $\nu_e^i$ for
every process $P_i$. In particular, $\forall \sigma_0 \in
\Sigma_0$,
\[
\nu_e^i \equiv (\sigma_0, \sigma') \in \Gamma_e \Leftrightarrow \left\{
\begin{array}{ll} 1.\, &
\forall x \in \U:\  \sigma'(x) = \sigma_0(x)\\
2.\, & \forall \tau \in \T:\ \sigma'(t) \leq \sigma'(\tau)\\
3.\, & \sigma_0(\pi_i) =\  \perp \mbox{ and } \sigma'(\pi_i) = l_0^i
\end{array} \right.
\]
{\sf Time Progress Transition: } The first kind of edges $\nu_+
\in \Gamma_+$ are those where the global clock is increased to the
minimum of all timeouts. In particular,
\[
\nu_+ \equiv (\sigma, \sigma') \in \Gamma_+ \Leftrightarrow \left\{ \begin{array}{ll}
1.\, &
\sigma(t) < \min\{\sigma(\T)\} \\
2.\, & \forall \tau \in \T:\  \sigma'(\tau) = \sigma(\tau) \\
3.\, & \forall x \in \X:\ \sigma'(x) = \sigma(x) \\
4.\, & \forall i:\ \sigma'(\pi_i) =\ \sigma(\pi_i)\\
5.\, & \sigma'(t) = \min\{\sigma(\T)\}
\end{array} \right.
\]
{\sf Timeout Increment Transition: } For the second kind of edges
$\nu_0^i \in \Gamma_0$ the global clock equals the minimum of
timeouts. Also if an edge in the TTD for process $P_i$
connects source location $l_j^i$ to
target location $l_k^i$ and is labeled by the instruction $\rho
\Rightarrow \lan \tau_i := update_i, \gamma, f \ran$, then
\[
\nu_0^i \equiv (\sigma, \sigma') \in  \Gamma_0 \Leftrightarrow \left\{
\begin{array}{ll} 1.\, &
\rho \mbox{ holds in } \sigma \\
2.\, & \sigma'(t) = \sigma(t) \\
3.\, &  \mathbf{If}\, \sigma(\tau_i) = \sigma(t)\\
& \ \ \ \mathbf{then }\, \sigma'(\tau_i) = update_i > \sigma(\tau_i)\\
& \mathbf{else}\, \sigma'(\tau_i) = \sigma(\tau_i)\\
4.\, & \forall x \in \gamma:\ \sigma'(x) = \sigma(t)\ \mbox{ and}\\
& \forall x \in \X\setminus\gamma:\ \sigma'(x) = \sigma(x) \\
5.\, & \forall v \in G \cup L_i: \sigma'(v) = f(\sigma(v))\ \mbox {and}\\
& \forall v \in \mathit{Var}\setminus(G \cup L_i):\ \sigma'(v) = \sigma(v)\\
6.\, & \sigma(\pi_i) =\ l_j^i \mbox{ and } \sigma'(\pi_i) = l_k^i
\end{array}
\right.
\]
If $update_i= k_1$ s.t. $l + z \prec k_1 \prec m + z'$, then $update_i$ arbitrarily selects a value $\delta$ such that $[l + \sigma(z) \prec \delta \prec m + \sigma(z')] \wedge [\delta > \sigma(\tau_i)]$ and returns $\delta$. If $update_i= k_2$ s.t. $k_2 \succ l + z$, then $update_i$ arbitrarily selects a value $\delta$ such that $[\delta \succ l + \sigma(z)] \wedge [\delta > \sigma(\tau_i)]$ and returns $\delta$. If $update_i= \infty$, $update_i$ returns the largest possible constant defined as per the design of the system. If $update_i= \max(\cM)$, $update_i$ nondeterministically selects any integer $\delta$ in $[0, M+1]$, where $M$ is the maximum of all the integers in $\cM$ returned by $\max(\cM)$. The local timing variables in $\gamma \subseteq \X_i$ for process $P_i$ are assigned the current value of global clock on timeout increment transition, while the other local timing variables in the system retain their old values before this transition. The variables in $\gamma$ are thus used to capture the delay between two events.\\ \\
{\sf Synchronous Communication:} For a pair of processes $P_s, P_r$ having synchronous communication edges $(e_s, e_r)$ as defined before,
$\nu_{syn\_comm}^{sr} \in \Gamma_{syn\_comm}$ exists such that:
\[
\nu_{syn\_comm}^{sr} \equiv (\sigma, \sigma') \in \Gamma_{syn\_comm} \Leftrightarrow \left\{
\begin{array}{ll} 1.\, &
\rho \mbox{ holds in } \sigma \\
2.\, & \sigma'(t) = \sigma(t) \\
3.\, &  \sigma'(\tau_s) = update_s > \sigma(\tau_s)\ \mbox {and}\ \\
&  \sigma'(\tau_r) = update_r > \sigma(\tau_r)\\
4.\, & \forall x \in (\gamma \cup \gamma'):\ \sigma'(x) = \sigma(t)\ \mbox{ and}\\
& \forall x \in \X\setminus(\gamma \cup \gamma'):\ \sigma'(x) = \sigma(x) \\
5.\, & \sigma'(\bar{m}) = \sigma(m)\\
6.\, & \forall v \in G \cup L_s: \sigma'(v) = g(\sigma(v))\\
& \forall v \in G \cup L_r: \sigma'(v) = h(\sigma(v))\\
& \forall v \in \mathit{Var}\setminus(G \cup L_s \cup L_r):\ \sigma'(v) =
\sigma(v)\\
7.\, & \sigma(\pi_s) =\ l_j^s, \sigma(\pi_r) =\ l_j^r \mbox{ and }\\
&  \sigma'(\pi_s) = l_k^s, \sigma'(\pi_r) =\ l_k^r
\end{array}
\right.
\]
This semantic model defines the set of possible computations of the ToM $P$ as a (possibly infinite) set of state sequences  $\xi:\,\, \sigma_0 \rightarrow \sigma_1 \rightarrow \ldots$, which starts with some initial state $\sigma_0$ in $\Sigma_0$ and follows with consecutive edges in $\Gamma$, {\it i.e.}, $\forall i. (\sigma_i, \sigma_{i+1}) \in \Gamma$. Let $\comp{S_P}$ be the set of all these computations of a ToM $P$ as defined by its TCTS $S_P$.

\subsection{Calendar Based Timed Transition Model}

\subsubsection{Syntax}
Next we capture bounded message transfer delay associated with an asynchronous communication.
Towards that the ToM is extended with a calendar data structure. A calendar is a linear
list of bounded size, where each element of the list contains the following information: {\sf message, sender\_id, receiver\_id, {\text and} expected\_delivery\_time}. Assuming $\C$ to denote the calendar array, a globally shared object, we set $$\U = \{t\} \cup \T \cup \X \cup \mathit{Var} \cup \C$$
Sending a message in a TTD of process $P_i$ is represented using the following edge:$$l_j^i\ \stackrel{\rho \Rightarrow \lan send(m, i,
\Omega), \tau_i := update_i, \gamma, f \ran}{\longrightarrow}\
l_k^i,$$ where $\Omega \subseteq R \times \Lambda$, $R \subseteq \{1,
2, \ldots n\}$ is the index set for the processes and  $\Lambda$ is the
set of expected message delays. $send(\ldots)$ specifies that a message $m$ is to be
sent to each of the processes $P_r$ with expected delivery time of $\lambda_r$
where $(r, \lambda_r) \in \Omega$. On taking a transition on this edge an
entry \{ {\sf m, i, r,} $\lambda_r$\} is added to $\C$ for each $(r, \lambda_r) \in \Omega$.

Receiving of the corresponding message is represented in the
TTD for each of the processes $P_r, \forall\ r \in R$ using the following edge:
$$l_j^r\  \stackrel{True \Rightarrow \lan receive(m, i, r), \tau_r:= update_r, \gamma, g \ran}{\longrightarrow}\ l_k^r,$$
where $receive(\ldots)$ specifies that a message $m$ sent by process
$P_i$ is to be received by the process $P_r$. On taking a transition on this
edge, the entry {\sf \{m, i, r, $\lambda_{r}$\}} is deleted from
$\C$.

\subsubsection{Semantics} Given a calendar $\C$, we assume that the set of delays for
all undelivered messages at any state $\sigma$ can be found using the function
$$\Delta: \sigma(\C) \rightarrow 2^{\N}$$
Again $\Gamma = \Gamma_e \cup \Gamma_+ \cup \Gamma_0 \cup
\Gamma_{syn\_comm} \cup \Gamma_{asyn\_comm}$ is the set of
transitions in the {\it calendar based clocked transition system} (CCTS).
Both $\Gamma_e$ (set of Entry Transition) and $\Gamma_{syn\_comm}$ (Synchronous Communication)
are same as in TCTS defined earlier.
The definitions for the edges in Time Progress Transition ($\Gamma_+$)
and those for Timeout Increment Transition ($\Gamma_0$) are
modified using calendar $\C$ as follows:\\\\
{\sf Time Progress Transition:} The first kind of edges $\nu_+$
are those where the global clock is increased to the minimum of all
the timeouts and message delays. In particular,
\[
\nu_+ \equiv (\sigma, \sigma') \in \Gamma_+ \Leftrightarrow \left\{
\begin{array}{ll} 1.\, &
\sigma(t) < \min\{\sigma(\T) \cup \Delta(\sigma(\C))\} \\
2.\, & \forall \tau \in \T:\  \sigma'(\tau) = \sigma(\tau) \\
3.\, & \forall x \in \X \cup \mathit{Var}:\ \sigma'(x) = \sigma(x) \\
4.\, & \forall i:\ \sigma'(\pi_i) =\ \sigma(\pi_i)\\
5.\, & \sigma'(t) = \min\{\sigma(\T) \cup \Delta(\sigma(\C))\}
\end{array} \right.
\]
{\sf Timeout Increment Transition:} For the second kind of edges
$\nu_0^i$ where global clock equals the minimum of all the timeouts
and message delays, we have: if an edge in the TTD of process $P_i$ connects source location
$l_j^i$ to target location $l_k^i$ and is labeled by the
instruction $\rho \Rightarrow \lan \tau_i := update_i, \gamma, f
\ran$, then
\[
\nu_0^i \equiv (\sigma, \sigma') \in \Gamma_0 \Leftrightarrow \left\{
\begin{array}{ll} 1.\, &
\rho \mbox{ holds in } \sigma \\
2.\, & \sigma'(t) = \sigma(t) \\
3.\, & \mathbf{If}\, [\sigma(t) = \min\{\sigma(\T)\}] \wedge [\sigma(\tau_i) = \sigma(t)]\\
& \ \ \ \mathbf{then }\, \sigma'(\tau_i) = update_i > \sigma(\tau_i)\\
& \mathbf{else }\, \sigma'(\tau_i) = \sigma(\tau_i)\\
4.\, & \forall x \in \gamma:\ \sigma'(x) = \sigma(t)\ \mbox{ and}\\
& \forall x \in \X\setminus\gamma:\ \sigma'(x) = \sigma(x) \\
5.\, & \forall v \in G \cup L_i: \sigma'(v) = f(\sigma(v))\ \mbox{ and}\\
& \forall v \in \mathit{Var}\setminus(G \cup L_i):\ \sigma'(v) = \sigma(v)\\
6.\, & \sigma(\pi_i) =\ l_j^i \mbox{ and } \sigma'(\pi_i) = l_k^i\\
\end{array}
\right.
\]
We additionally define new transitions corresponding to $send()$
and $receive()$ to capture asynchronous communication:\\\\
{\sf Send Transition:} If there is an edge in process $P_i$, which
connects source location $l_j^i$ to target location $l_k^i$ and is
labeled by the instruction $\rho \Rightarrow \lan send(m, i, \Omega), \tau_i := update_i, \gamma, f \ran$, then we have the corresponding edge $\nu_{send}^i \in \Gamma_{asyn\_comm}$, which
adds $|\Omega|$ cells to the calendar array $\C$:
\[
\nu_{send}^i \equiv (\sigma, \sigma') \Leftrightarrow \left\{
\begin{array}{ll} 1.\, &
\rho \mbox{ holds in } \sigma \\
2.\, & \sigma'(t) = \sigma(t) \\
3.\, & \sigma'(\tau_i) = update_i > \sigma(\tau_i)\\
4.\, & \forall x \in \gamma:\ \sigma'(x) = \sigma(t)\ \mbox{ and}\\
& \forall x \in \X\setminus\gamma:\ \sigma'(x) = \sigma(x) \\
5.\, & \forall v \in G \cup L_i: \sigma'(v) = f(\sigma(v))\ \mbox{ and}\\
& \forall v \in \mathit{Var}\setminus(G \cup L_i):\ \sigma'(v) = \sigma(v)\\
6.\, & \forall (r, \lambda_r) \in \Omega: \sigma'(\C) := \sigma(\C) \cup \mbox{{\sf \{m, i, r, $\lambda_r$\}}}\\
7.\, & \sigma(\pi_i) =\ l_j^i \mbox{ and } \sigma'(\pi_i) = l_k^i
\end{array}
\right.
\]
{\sf Receive Transition:} If there is an edge in process $P_r$,
which connects source location $l_j^r$ to target location $l_k^r$
and is labeled by the instruction $True\Rightarrow \lan receive(m,
i, r), \tau_r := update_r, \gamma, g \ran$, then we have the
corresponding edge $\nu_{receive}^r \in \Gamma_{asyn\_comm}$,
which deletes the entry $\{m, i, r, \lambda_r\}$ from the calendar
array $\C$ when the clock $t$ reaches $\lambda_r$:
\[
\nu_{receive}^r \equiv (\sigma, \sigma')  \Leftrightarrow \left\{
\begin{array}{ll} 1.\, &
\exists \{m, i, r, \lambda_r\} \in \sigma(\C)\ \mbox{s.t. } \sigma(t) = \lambda_r\\
2.\, & \sigma'(t) = \sigma(t) \\
3.\, & \sigma'(\tau_r) = update_r > \sigma(\tau_r)\\
4.\, & \forall x \in \gamma:\ \sigma'(x) = \sigma(t)\ \mbox{ and}\\
& \forall x \in \X\setminus\gamma:\ \sigma'(x) = \sigma(x) \\
5.\, & \forall v \in G \cup L_r: \sigma'(v) = g(\sigma(v))\ \mbox{ and}\\
& \forall v \in \mathit{Var}\setminus(G \cup L_r):\ \sigma'(v) = \sigma(v)\\
6.\, & \sigma'(\C) := \sigma(\C)\setminus\{m, i, r, \lambda_r\}\\
7.\, & \sigma(\pi_r) =\ l_j^r \mbox{ and } \sigma'(\pi_r) = l_k^r\\
\end{array}
\right.
\]
Similar to the case of TCTS, this semantic model also defines the set of possible computations of the calendar based ToM as a (possibly
infinite) set of state sequences starting with some initial state in $\Sigma_0$ and following consecutive edges in $\Gamma$.
Let $\comp{S_P}$ be the set of all these computations of a calendar based ToM $P$ as defined by its CCTS $S_P$.\\\\
{\bf Models for Time:} It remained unspecified as to what would be the underlying model of time for clock, timeouts etc that appear in the definitions of TCTS and CCTS. There are two natural choices for time, the set of non-negative integers $\N$ (discrete time) or the set of non-negative reals $\mbR$ (dense time). Given the model of time as $\mathit{TIME}$, let $\comp{S_P}_{\mathit{TIME}}$ be the set of all the computations of a ToM (or calendar based ToM) $P$ as defined by its TCTS (or CCTS) $S_P$.

When we consider that the underlying model of time as $\mbR$, we need to add the following non-zenoness condition to ensure effective time progress in the model. {\it There must not be infinitely many time progress (or timeout increment) transitions effective within a finite interval.}
Formally,\\\\
    {\sf nonzenoness:} $$\forall \xi:\,\, \sigma_0 \rightarrow \sigma_1 \rightarrow \ldots \in \comp{S_P}_{\mbR}.\forall \delta \in \mbR. \exists i\geq 0. \sigma_{i}(t) > \delta$$ 

\subsection{Parametric Processes}

We consider the case of finite family of processes specified in a parametric way. A completely parametric process family would be specified as $$\{\theta\}[\{P(i)\}_{i = 1}^{i=N}]$$ where $N \geq 1$ is some finite positive integer and  $\theta= \theta_1 \wedge \ldots \wedge \theta_N$ such that $\theta_i$ (${1\leq i\leq N}$) initializes the variables for the $i^{th}$ copy of the process. Process $P(i)$ could be a TTD or a calender based TTD. 

The semantic interpretation of such parametrically specified process family is given by first flattening the specification as $$\{\theta\}[P(1) || \ldots || P(N)]$$ and then applying the semantics presented before as per the case of $P(i)$ being a TTD or a calendar based TTD. 

Such parametric specification can be generalized to a homogeneous set of process families as $$\{\theta\}[ \{P(i_1)\}_{i_1 = 1}^{i_1 = N_1}|| \ldots || \{P(i_l)\}_{i_l = 1}^{i_l = N_l} ]$$ where $N_1, \ldots N_l$ are some finite positive integers and $\theta= \mathbf{\theta_1 \wedge \ldots \wedge \theta_l}$ such that $\mathbf{\theta_i} = \theta_{i1} \wedge \ldots \wedge \theta_{iN_i}$ initializes the variables for the $i^{th}$ process family. The term homogeneous arises because processes  in all the process families should uniformly be either TTDs or calender based TTDs. We do not consider the case of hetrogeneous set of process families, where processes across different process families might be different. Similar to the case of a single parametric process family, the generalized process family can be interpreted by flattening the process specification.  

\section{Examples}

Following two examples would illustrate the expressiveness and effectiveness of the proposed timeout and calendar based modeling framework.

\subsection{Fisher's Mutual Exclusion Protocol}

\begin{figure}
\centering
\includegraphics[width=5.0in]{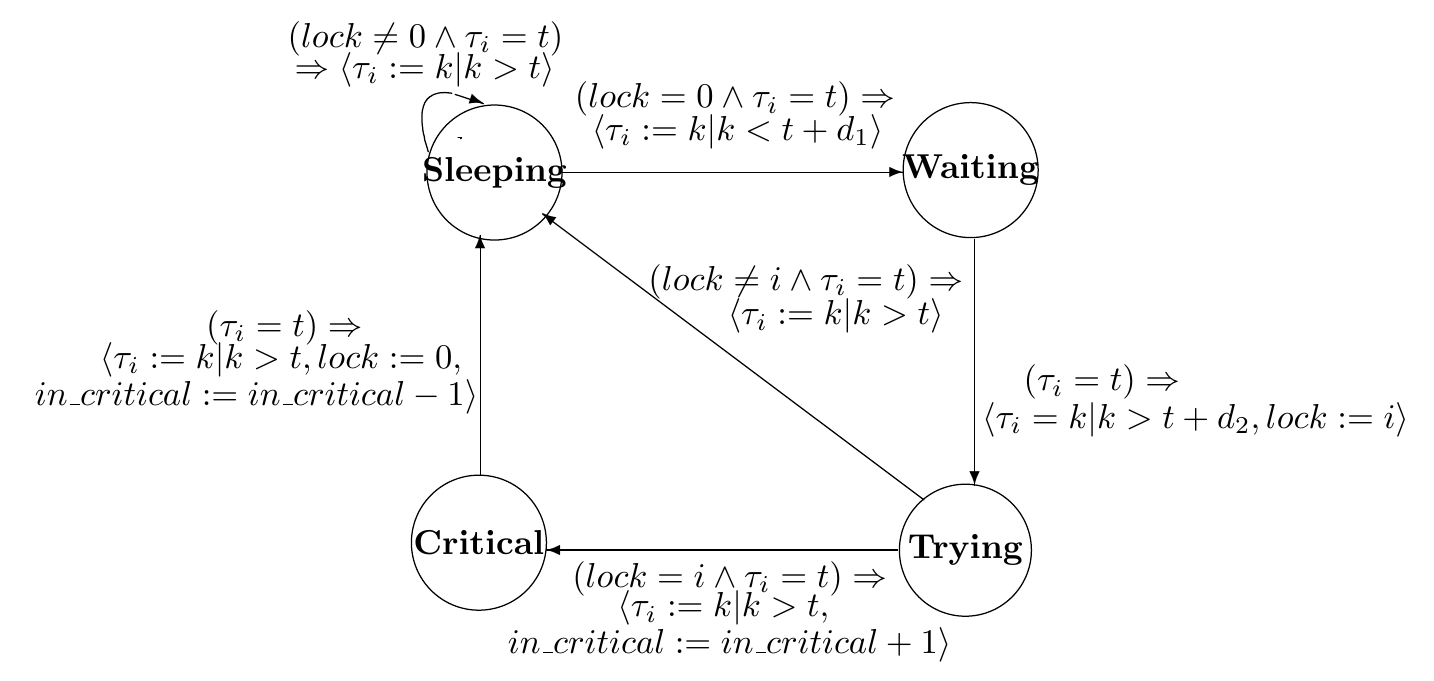}
\caption{TTD for the $i^{th}$ process in the Fischer's Protocol} \label{fig:fig2}
\end{figure}

Fischer's protocol is a well studied protocol to ensure mutual exclusion
among real time concurrent processes. Let there be $n$ processes
$P_1,\ldots, P_n$ trying to access shared resources in a real-time fashion to be discussed
later. A process $P_i$ is initially idle (\emph{Sleeping} state), but at
any time, may begin executing the protocol provided the value of a global
variable $lock$ is $0$ and then move to \emph{Wait} state. There it can wait
up to maximum of $d_1$ time units before assigning the value $i$ to $lock$ and moving to
\emph{Trying} state. It may enter the {\em Critical} section after a delay of at least of $d_2$
time units provided the value of $lock$ is still $i$. Otherwise it
has to move to \emph{Sleeping} state. Upon leaving the {\em Critical}
section, it re-initializes $lock$ to $0$. There is another global
variable, $in\_critical$, used to keep count of the number of
processes in the critical section. The auto-increment
(auto-decrement) of the variable is done before a process enters
the {\em Critical} section (leaves the {\em Critical} section). Mutual exclusion is ensured if $d_1 < d_2$. The
timeout-based TTD of the $i^{th}$ process
$P_i$ executing Fischer's protocol is shown in Figure~\ref{fig:fig2}.

\subsection{TTA Startup Algorithm}\label{clocked-tta}
\begin{figure}
\centering
\includegraphics[scale=0.7]{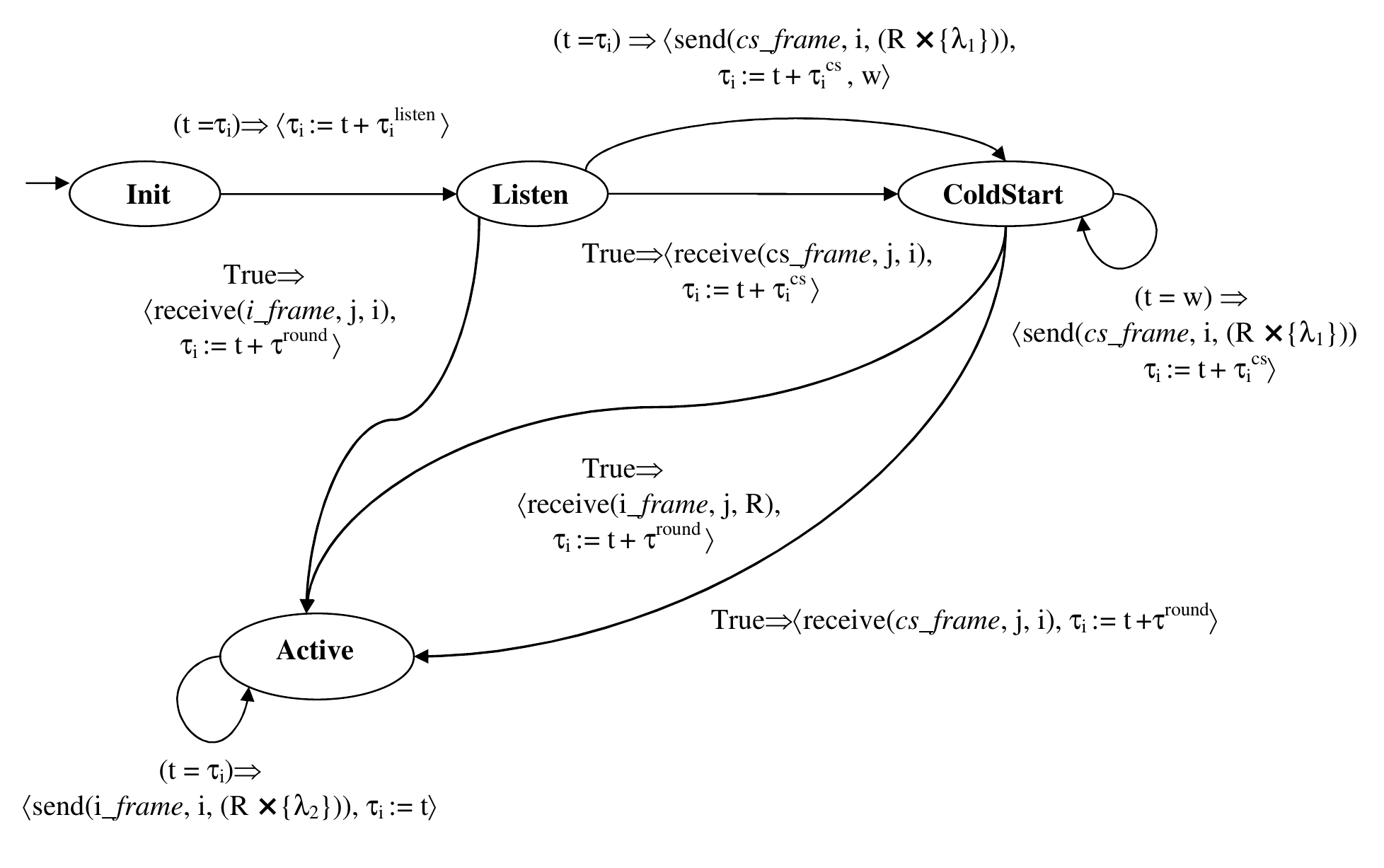}
\caption{Calendar-based TTD for the $i^{th}$ node in TTA Startup algorithm } \label{fig:fig3}
\end{figure}
The TTA startup algorithm can be formalized using the calendar based model described above.
This algorithm executes on a logical bus meant for safety-critical
application in both automotive and aerospace industries. In a normal
operation, $N$ processors or nodes share a TTA bus using a
TDMA schedule. The goal of the startup algorithm
is to bring the system from the power-up state, in which the
processors are not synchronized, to the normal operation mode in which
all processors are synchronized and follow the same TDMA schedule.

In TTA startup algorithm each node $i \in
\{1\ldots N\}$ has two unique timeout parameters, $\tau_i^{listen}$
and $\tau_i^{cs}$, for $listen$ and $coldstart$ states respectively. These are defined as follows:
\[
\begin{array}{l}
\hspace{5cm}\tau_i^{listen} = 2\tau^{round} + \tau_i^{startup},\\
\hspace{5cm}\tau_i^{cs} = \tau^{round} + \tau_i^{startup}
\end{array}
\]
where $\tau^{round}$ represents the TDMA round duration and
$\tau_i^{startup}$ denotes the duration between the start of a
TDMA cycle and the time when the slot for node $i$ starts. If
$\tau$ denotes the duration of each slot then
$$\tau^{round} = N\tau ,\ \  \tau_i^{startup} = (i-1)\tau$$
When a node is powered-on, it performs some internal
initialization, and transits to the \emph{Listen} state. In this state it listens for
the unique duration $\tau_i^{listen}$ to determine if there is a
synchronous set of nodes communicating on the medium. The nodes
which are in the \emph{Active} state are already synchronized, and periodically transmit i-frames that
carry the TDMA cycle structure. If a node in the \emph{Listen}
state receives such an i-frame, it adjusts its state to the frame
contents and is thus synchronized with the set of already synchronous
nodes. If the above does not happen, there are two possibilities.
Each node listens for a "cold-start" message (cs-frame) from
another node indicating the beginning of the cold-start sequence;
cs-frames are similar to i-frames but carry a protocol state
suggested by the sending node. When a node completes the reception of
a cs-frame, it enters the \emph{Coldstart} state and resets its
local clock to $\delta_{cs}$ (that is the transmission duration of
the cs-frame). Thus, all nodes that received the cs-frame have
synchronized local clocks (within system tolerances, including the
propagation delay). Each node that receives neither an i-frame nor
a cs-frame during the \emph{Listen} phase enters
the \emph{Coldstart} state on its listen timeout, resets its local clock to 0 and broadcasts a
cs-frame. Thus, after the transmission of the cs-frame
($\delta_{cs}$ later), the local clock of the sending node is
also synchronized to the local clocks of the set of receiving
nodes.

Each node in the \emph{coldstart} state waits for reception of
another cs-frame or i-frame until its local clock reaches the
value of its individual cold-start timeout. If it
receives such a frame it synchronizes on its contents and enters
the \emph{Active} state; if not, it resets its local clock and again
broadcasts a cs-frame. No further collision can occur at this
point, because cold-start timeouts have a strict
order and that is why no two nodes that caused a collision can
collide again. The listen timeout of any node is greater than
coldstart timeout of any node. No node which has come in the
\emph{Listen} state after the collision cannot move to the
\emph{Coldstart} state before the collision is resolved. For further details
of startup protocol, we refer the reader to~\cite{SP02}.

The calendar based TTD of the $i^{th}$ node
is depicted in Figure~\ref{fig:fig3}. In TTA startup algorithm,
all the communications are asynchronous and hence, message delivery delays,
which are finite and specified by the designer have to be taken into
account for correct operation of the protocol. The timeouts $\tau_i^{listen}$ and
$\tau_i^{cs}$ represent how much time a node spends in
\emph{Listen} state and \emph{Coldstart} state respectively, if no
external signal is received. The timeout $\tau^{round}$ denotes the time a
node spends in \emph{Active} state before sending its next massage.
$R_i = \{1, \ldots, N\}\setminus\{i\}$ represents the set of nodes except
the sender $i$ that are required to receive the message in the network. We use
$\lambda_i$'s to denote the message delivery time for the corresponding
send events. In TTA, message delivery times for all the receivers
are considered to be the same, and that is why we have considered a
single variable $\lambda_i$ to represent that delay.

\section{Verification Results for Digitization} \label{digitize}

In literature the verification problem for real-time systems assumes two
descriptions of real-time behavior, implementation $I$ and specification $S$, and poses the question whether
$I$ implements/satisfies $S$. The implementation language ${\cal L}_{I}$ describes systems and behavior
over time while the specification language ${\cal L}_{S}$ describes the timing requirements of the system. The verification
obligation involves presenting algorithms and/or proof rules that facilitate a formal argument that a particular implementation meets the requirement of a particular system under some particular assumption of semantics of computation and time. Assuming $C$ and $T$ to be mathematical models of computation and time respectively, the real-time verification problem parameterized
by $(C, T, {\cal L}_{I}, {\cal L}_{S})$ states: does the implementation of the system $I$, given as an expression of ${\cal L}_I$
meet the specification $\phi$ given as an expression of ${\cal L}_S$, with respect to the semantical assumption $(C,T)$, written as  $$I \models_{(C,T)}^{?} \phi$$
In particular, we would consider two important instances of the real-time verification problem - one with an {\it integral} model of time and one with a {\it dense} model of time. In the following, we assume TTS as the implementation language and linear time temporal logic (LTL) as the specification formalism.

\subsection{Timed Sequences}
We shall adopt discrete trace model (using the terminology from~\cite{HMP92,Bos99})
as a mathematical model of computation. By {\em discrete trace} model one can capture
the behavior of a system as an infinite sequence of snapshots of the
global system state at certain times. We assume our time domain $TIME$ has
a total ordering $\leq$ defined on it. We define an {\em observation} to be a pair $(\sigma_i, T_i)$,
where $\sigma_i$ is a state and $T_i \in TIME$. A {\em timed state sequence} $\eta = (\sigma, T)$
is an infinite sequence $\eta: \, \, (\sigma_0, T_0) \rightarrow (\sigma_1, T_1) \rightarrow (\sigma_2, T_2) \rightarrow \cdots$ of observations\footnote{Note that any  $\xi \in \comp{S_P}$ (previously defined) essentially defines a timed state sequence. This is because, states in $\xi$ have implicit representation for time stamps as $\sigma_0(t), \sigma_1(t), \ldots$, which are otherwise explicitly present in the definition of $\eta$ as $T_0, T_1, \ldots$}. Further, the infinite sequence $T_i \in T$ of time stamps in $\eta$ satisfy (i) {\em monotonicity}: $T_i \leq T_{i+1}$ for all $i \geq 0$, and (ii) {\em progress:} time progresses, for all $T \in TIME$, $T_i \geq T$ for some $i \geq 0$.

Now onwards, we shall work with dense-time models when $TIME = \mbR$ and integral-time models when $TIME = \mbN$. A timed state sequence under dense-time model will be referred to as {\em precisely timed} and under integral-time model as  {\em digitally timed}.

Let us denote the set of all timed state sequences over the $TIME$ domain as
$TSS_{\mathit{TIME}}$. A {\it real-time} property is a subset of $TSS_{TIME}$. 
Every real-time system $S$ defines a real-time property, denoted as $\comp{S}$, which is the set of all timed state sequences of $S$. Also, every real-time specification $\phi$ defines a real-time property $\comp{\phi}$,
the set of real-time sequences that satisfy $\phi$.

Now let us formulate the real-time verification problem. We say a real-time system $S$
satisfies the specification $\phi$, written as $$ S \models_{TIME} \phi$$ if and only if
$${\comp{S}}_{TIME} \subseteq {\comp{\phi}}_{TIME} $$

Consider a dense-time property $\Pi_{\mbR} \subseteq {TSS}_{\mbR}$, a set of
of timed state sequences over $\mbR$. Its {\it clock-independent semantics}
${\mbN}(\Pi_{\mbR})$ is the subset of digitally timed state sequences in
$\Pi_{\mbR}$, {\em i.e.,} ${\mbN}(\Pi_{\mbR}) = \Pi_{\mbR} \cap {TSS}_{\mbN}$.
In~\cite{HMP92}, it is shown that clock-independent semantics is not very adequate for reasoning about dense time. As a remedy of this, another approximate semantics was
introduced, which was called {\it digitization}.

The following definitions will be useful for our subsequent discussions. %
For any timed state sequence $\eta = (\sigma, T)$, we introduce {it untime} operation
$\eta^{-}$ as its state component $\sigma$. Also, $\eta^{i} = (\sigma^{i}, T^{i})$, for $i \geq 0$, denotes the timed state sequence that results from $\eta$ by deleting the first $i$ observations (note, $\eta^{0} = \eta$).

\subsection{Digitization}
Given $x \in \mbR$ and $\ep \in (0,1]$, we define $\dig{x}_{\ep} = \lfloor x \rfloor$ if
$x \leq \lfloor x \rfloor + \ep$, otherwise $\dig{x}_{\ep} =\lceil x \rceil$\footnote{where
$\lfloor{\cdot}\rfloor$ and $\lceil{\cdot}\rceil$ are the floor and ceiling rounding operations
on real numbers respectively}. Given a precisely timed sequence
$\eta = (\sigma, T)$ and $\ep \in (0,1]$, we define the $\ep$-{\em digitization} $\dig{\eta}_{\ep} = (\sigma,
\dig{T}_{\ep})$ of $\eta$ be the digitally timed sequence
\[ (\sigma_0, \dig{T_0}_{\ep}) \rightarrow (\sigma_1, \dig{T_1}_{\ep}) \rightarrow \cdots, \]
For any dense-time property $\Pi$ (a set of timed sequences over dense time) let
$$\dig{\Pi} = \{\dig{\eta}_{\ep} \,|\, \eta \in \Pi \,\, \mbox{and} \,\, \ep \in (0,1] \},$$ which
is a digitization of $\Pi$. We write $\dig{\eta}$ instead of $\dig{\{\eta\}}$.

We state some concepts from~\cite{HMP92}. Let $\Pi$ be a dense-time property.
$\Pi$ is {\em closed under digitization} iff for all $\eta \in {TSS}_{\mbR}, \eta \in \Pi$ implies
$\dig{\eta} \subseteq \Pi$. $\Pi$ is {\em closed under inverse digitization} iff $\dig{\eta}
\subseteq \Pi$ implies $\eta \in \Pi$, for all $\eta \in {TSS}_{\mbR}$. Finally, $\Pi$ is
{\em digitizable} iff it is closed under both digitization and inverse digitization, {\it i.e.,}
$\eta \in \Pi$ iff $\dig{\eta} \subseteq \Pi$ for all $\eta \in {TSS}_{\mbR}$. We state the
following important result (see~\cite{HMP92}).
\begin{Fact}
Assume a real-time system $S$ whose dense-time semantics $\comp{S}_{\mbR}$ is closed under digitization, and a specification $\phi$ whose dense-time semantics ${\phi}_{\mbR}$ is closed under inverse digitization. Then in
order to prove $S \models_{\mbR} \phi$ it suffices to check if $S \models_{\mbN} \phi$.
\end{Fact}
A dense-time property $\Pi$ is said to be {\em qualitative} if $\eta \in \Pi$ implies $\eta' \in \Pi$
for all precisely timed sequences $\eta$ and $\eta'$ with identical state components ({\it i.e.,}
$\eta^- = \eta'^{-}$).
\begin{Fact}~\cite{HMP92} Every qualitative property is digitizable.
\end{Fact}

\subsection{Digitization of Timeout and Calendar based Transition Systems}~\label{digi_tom}
Recall a TCTS is $S = (\V, \Sigma, \Sigma_0, \Gamma)$ (we drop the subscript $P$ because we assume the ToM $P$ is implicit) where $\V$ is a set of variables, $\Sigma$ a set of states, ${\Sigma}_0 \subseteq \Sigma$ a set of initial states and $\Gamma$ a set of transitions. We would like to show that the computations for this transition system
are digitizable. Our approach follows~\cite{Bos99}.

A {\it run} of $S$ over a timed state sequence $\eta : (\sigma_0, T_0) \ra (\sigma_1, T_1) \ra \cdots$ is
a  sequence of pairs of $S$ of the form $\zeta: (\sigma_0, \nu_0) \stackrel{T_1}{\ra} (\sigma_1, \nu_1) \stackrel{T_2}{\ra} \cdots$ where $\sigma_i$ denotes the state and $\nu_i$ the mapping of variables in $\U$ in state $\sigma_i$ and further, it satisfies the following conditions:
\begin{enumerate}
          \item (initiation:) $\sigma_0 \in \Sigma_0$ and $\nu_0(t) = T_0,\;\; \forall i \geq 0.\nu_0(\pi_i) = \perp,\;\; t \in \V,  \pi_i \in \V$.
          \item (consecution:) for $i \geq 1$ there is an edge $({\sigma}_{i-1}, \sigma_i) \in \Gamma = (\Gamma_e \cup \Gamma_+ \cup \Gamma_0 \cup \Gamma_{syn\_comm})$ such that the following hold:
              \begin{itemize}
                \item if $(\sigma_0, \sigma_1) \in \Gamma_e$ then $T_{0} = \nu_0(t) \leq \nu_1(t) = T_1$ and $\forall \tau \in \T.\, \sigma_1(\tau) \geq T_1$.
                \item if $(\sigma_{i-1}, \sigma_i) \in \Gamma_+$ then $T_{i-1} = \nu_{i-1}(t) < \min\{\sigma_{i-1}(\T)\} = \nu_i(t) = T_i$.
                \item if $(\sigma_{i-1}, \sigma_i) \in \Gamma_0$ then $T_{i-1} = \nu_{i-1}(t) = \nu_i(t) = T_i$.
                \item if $(\sigma_{i-1}, \sigma_i) \in \Gamma_{syn\_comm}$ then $T_{i-1} = \nu_{i-1}(t) = \nu_i(t) = T_i$.
              \end{itemize}
          \item (time progress:) for any real number $T$ there exists an $i \geq 0$ such that $T_i > T$.
\end{enumerate}

We say that $\eta \in {TSS}_{\rm TIME}$ is {\em time-consistent} (for $S$) if $S$ has a run over it. In the sequel we consider only time-consistent behaviors $\eta \in \comp{S}_{\rm TIME}$ of $S$, {\it i.e.,} $\eta \in \comp{S}_{\rm TIME}$ iff there is run over $\eta$. If ${\rm TIME} = \N$ then we get integral behavior of TCTS. Now it is obvious that time at state $j \geq 1$ in a given run, is given by $\nu_j(t) = T_j$. We define $\ep$-digitization of the mapping $\nu_j$ for any variable $x \in \U \subseteq \V$ as ${\lan\nu_j(x)\ran}_{\ep} = [\nu(x)]_{\ep}$.

Given a computation $\zeta: (\sigma_0, \nu_0) \stackrel{T_1}{\ra} (\sigma_1, \nu_1) \stackrel{T_2}{\ra} \cdots$ its $\ep$-digitization is the computation ${[\zeta]}_{\ep}: ({\sigma_0, \lan \nu_0 \ran})_{\ep}) \stackrel{[T_1]_{\ep}}{\ra} ({\sigma_1, \lan \nu_1\ran})_{\ep}  \stackrel{[T_2]_{\ep}}{\ra} \cdots$, where ${\lan {\nu_j} \ran}_{\ep}$ for $j \geq 1$ are defined above, and ${\lan \nu_0 (t)\ran}_{\ep} = [T_0]_{\ep}$.\\
\\
Now we need to analyze the extent to which the set of dense-time computations of a TCTS are closed under digitization.
Suppose $\zeta : (\sigma_0, \nu_0) \stackrel{T_1}{\ra} (\sigma_1, \nu_1) \stackrel{T_2}{\ra} \cdots$ is a run of $S$ over $\eta$. For digitization, ${[\zeta]}_{\ep}$ would be a run of $S$ over ${[\eta]}_{\ep}$. We have ${\lan \nu_0 (t)\ran}_{\ep} = [T_0]_{\ep}$. Observe if $T_{i-1} = T_i$ then ${[T_{i-1}]}_{\ep} = {[T_{i}]}_{\ep}$. When $T_{i-1} < T_i$, except for the case of $0 < (T_i - T_{i-1}) < 1$, we have ${[T_{i-1}]}_{\ep} < {[T_{i}]}_{\ep}$. So, if there is an edge $(\sigma_{i-1}, \sigma_i) \in \Gamma$ and $(T_i = T_{i-1}) \vee (T_i \geq T_{i-1} + 1)$, there would be an edge $({\lan \sigma_{i-1}\ran}_{\ep}, {\lan \sigma_i \ran}_{\ep})$ in $\Gamma$ under ${[\zeta]}_{\ep}$. Also we can ensure time progress for $[\zeta]_{\ep}$. Hence:

\begin{Fact}
The set of dense-time computations of a TCTS are closed under digitization if and only if all timeout increments are at least $1$ time unit.
\end{Fact}

The result above indicates a precise characterization for the digitization for a TCTS. All timeout increments in $(0,1)$ result into a TCTS, which are not closed under digitization and therefore cannot be model checked for all LTL properties under discrete time dynamics.

A similar argument can be used to show that the dense computations of a (digitizable) calendar based clocked transition system (CCTS) are also closed under digitization.

\subsection{Linear Temporal Logic}
Let us briefly describe propositional linear temporal logic~\cite{Pn77}, more popularly known as LTL. The vocabulary
of LTL consist of a set ${\cal P}$ of atomic propositions. The formulas of LTL are built using
boolean connectives, next operator $\bigc$ and {\em until} operator $\U$ as follows:
\[ \phi \,\,::= \,\, p \,|\, \neg \phi \, |  \, \phi_1 \wedge \phi_2 \, | \, \bigc \phi  \,| \, \phi_1 \U \phi_2, \;\;\;\;  p \in {\cal P} \]
The other temporal operators can be introduced as abbreviations, {\em e.g.}, $F \phi\;\; \define \;\; True\ \U \ \phi, G \phi \;\;\define \;\; \neg F \neg \phi$.

The formulas of LTL can be interpreted over timed state sequences whose states are from $\Sigma$
such that each state in $\Sigma$ gives rise to an interpretation for propositions in ${\cal P}$. Let
$\eta = (\sigma, T)$ be a timed state sequence with $\sigma_i \in \Sigma$ for $i \geq 0$. The
satisfaction relation $\eta \models \phi$ is defined inductively as follows:
\[
\begin{array}{ccc}
\eta \models p \,\, & {\rm iff}& \,\, \sigma_0 \models p; \\
\eta \models \neg \phi \,\, & {\rm iff} & \,\, \eta \not \models \phi; \\
\eta \models \phi_1 \wedge \phi_2 \,\, &{\rm iff}& \,\, \eta \models \phi_1 \,\,{\rm and} \,\, \eta \models \phi_2 \\
\eta \models \bigc \phi \,\, & {\rm iff} & \,\, \eta^1 \models \phi \,\, {\rm and} \,\, T_1 \geq T_0,\\
\eta \models \phi_1 \U \phi_2 \,\,  & {\rm iff} & \exists i \geq 0. \exists \alpha \in \N. {\eta}^i \models \phi_2, \,\, {\rm where} \\
\ &\ & \ \ \ \ \ \ \ \ \ \ \ T_i \geq T_0 + \alpha,  \,\, {\rm and} \,\, \forall j. 0 \leq j < i. {\eta}^j \models \phi_1.
\end{array}
\]
For a LTL-formula $\phi$, let the set ${\comp{\phi}}_{TIME} \subseteq {TSS}_{\rm TIME}$ contain all timed state sequences $\eta$ over the time domain $TIME$ such that $\eta \models \phi$. Thus, ${\comp{\phi}}_{\mbR}$ is the analog dense-time property for the formula $\phi$. Note that for any specification $\phi$ expressed in LTL, ${\comp{\phi}}_{\mbR}$ is closed under inverse digitization. To see this consider two timed sequences $\eta$ and $\eta'$ with identical state components. Suppose $\eta \models \phi$, {\it i.e.}, $\eta \in {\comp{\phi}}_{\mbR}$. Now the proof is by induction on the structure of $\phi$. At the induction stage, we only consider the case $\phi = \phi_1 \U \phi_2$. Now $\eta \models \phi_1 \U \phi_2 \,\, \mbox{iff for some} \,\, i \geq 0, \alpha \in \N, {\eta}^i \models \phi_2, \,\, {\rm where} \,\, T_i \geq T_0 + \alpha,  \,\, {\rm and} \,\, {\eta}^j \models \phi_1 \,\,\mbox{for all} \,\, 0 \leq j < i$. By induction hypothesis, we have $\eta'^i \models \phi_2$ and $\eta'^j \models \phi_1$. Since, $T'_i \geq T'_0$, there exists some $\alpha' \in \N$ such that $T'_i \geq T'_0 + \alpha'$. Therefore $\eta' \models \phi$ and hence $\eta' \in {\comp{\phi}}_{\mbR}$.

\subsection{An Integral Verification Problem}

We conclude this section with this important observation. Given a TCTS or CCTS $S$, corresponding to a timeout-based or a calender-based model and a specification formula $\phi$ in LTL we may check $S \models_{\mbR} \phi$ by verifying whether $S \models_{\mbN} \phi$. In the next section we shall try to further simplify this problem.

%
%


\section{Clockless Modeling}\label{clockless}

A finite state model-checker like {\sf Spin}~\cite{Hol03} uses finite state automata to model the behavior of concurrent processes in distributed systems. The
combined execution of a system of asynchronous processes is described as a product of automata each of which models an individual process.
The product automaton is finite if the number of processes, message channels, number of messages in a channel, and the range of values for various variables are finite in the automaton for each individual process. 

Though timeout and calendar based models can be used to efficiently capture dense time semantics without using a continuously varying clock, it is difficult to use these
models for finite state model checking, even though we have seen that in most of the cases the verification problem reduces to an integral one thanks to digitization. The difficulty arises from the fact that the value of the global clock $t$ and the values of the timeout variables in $\T$ diverge and thus are not bounded by a finite domain. Unlike TA there is no provision of resetting the global clock or timeouts in these models, as a result of which the timeout and calendar based models cannot be directly used for finite state model checking.

We propose a finitary reduction technique, which is formalized in terms of clockless modeling  and semantics in the next section. This
technique effectively reduces the timeout and calendar based
transition systems with discrete dynamics into finite state
systems, which, in turn, can be expressed and model checked by
finite state model checkers. The assumption of discrete time as the
underlying model is particularly relevant to cases where we are left with
integral verification problem exploiting digitization results.

From the semantics of the timeout based systems it is clear that
to implement time progress transition, a special process
is required to increase the global clock to the minimum of
timeouts, when each of the timeout values is strictly greater than
the current value of the clock. A process $P_i$ waits until its
timeout is equal to global clock, and when it is so, $P_i$ takes the
discrete transition and updates its own timeout according to the specified
updation rule. We model this special process, which is
responsible for time progress transition in such a way that it
does not explicitly use the clock variable and prevents the
timeout variables from growing infinitely. We call this process as
\emph{time\_progress}.

The process \emph{time\_progress} is implemented as follows.
When the global clock is less than all
the timeouts no discrete transition is possible in the system.
In such a situation,
\emph{time\_progress} finds out the minimum of all the timeouts in
$\T$ and scales down all these timeouts in $\T$ by this amount. In
this way at least one of the timeouts becomes zero. The guards of
the processes are defined in such a way that the processes wait
until their timeouts become zero. When it happens the process
updates its timeout and does other necessary jobs.



If $update$ function always increments the timeouts by a finite
value then it is guaranteed that the value of a timeout will
always be in a finite domain. But in some cases it is possible
that a timeout may take any value in the future. In those cases,
the value of the timeout is taken as the largest
possible value defined by the system. This approach can also be extended for the
calendar based models as well.

The discussion above is formalized in terms of ``clockless'' modeling as below:


\subsection{Timeout based Models: Clockless Modeling} \label{toclockless}

\subsubsection{Clockless Syntax} In order to capture the effect of finite state reduction in a timeout model, we restrict the set $\U$ and redefine $update_i$ as follows: $$\U = \T \cup \X \cup \mathit{Var}.$$ $update^-_i$ is given by the following rule: $$update^-_i = k_1\ |\ k_2\ |\ \infty |\ \max(\cM),$$ where $l - z \prec k_1 \prec' m - z'$ for $\prec,\prec' \in \{<, \leq\}$ and $k_2 \succ l - z$ for $\succ \in \{>, \geq\}$; $z, z' := w | 0$ and $l, m \in \N$ are non negative integer constants. For any $z \in U$ let $\sigma^-_i(z)$ stand for the value of the variable $z$ in (clockless) state $\sigma^-_i$. Note that $update^-_i$ is different from the update function $update_i$ for clocked transition system in the sense that this one updates the timeouts in bounded domain.

\subsubsection{Clockless Semantics} For clockless modeling of timeout based models we associate a transition system $S^-_P = (\V^-, \Sigma^-, \Sigma^-_0, \Gamma^-)$, where $\V^- = \V \setminus \{t\}$ is a set of variables, $\Sigma^-$ a set of clockless states, $\Sigma^-_0 \subseteq \Sigma^-$ initial clockless states (defined in an analogous manner as for clocked transition systems) and $\Gamma^-$ a set of clockless transitions. We remark that given a timeout based model, the set of states $\Sigma$ for clocked transition system and the set of states $\Sigma^-$ for clockless transition system are exactly similar modulo the assignment of the global clock variable $t$. The same is true for initial states too. Note $\Gamma^- = \Gamma^-_e \cup \Gamma^-_+ \cup \Gamma^-_0 \cup \Gamma^-_{syn\_comm}$, while $\Gamma^-_e$ is identical to $\Gamma_e$ for clocked transitions, we shall only define Time Progress Transition $\Gamma^-_+$, Timeout Increment Transition $\Gamma^-_0$, and Synchronous Communication Transition $\Gamma^-_{syn\_comm}$ by modifying the same for the clocked timeout transition system as defined earlier.\\ \\
{\sf Time Progress Transition:} The edges $\nu_+$ are redefined such that all the timeouts are decremented by the minimum of the current timeout values. In particular,
\[
\nu_+ \equiv (\sigma^-, \sigma'^-) \in \Gamma^-_+ \Leftrightarrow \left\{ \begin{array}{ll}
1.\, & \min\{\sigma^-(\T)\} > 0 \\
2.\, & \forall \tau \in \T:\  \sigma'^-(\tau) = \sigma^-(\tau) - \min\{\sigma^-(\T)\}\\
3.\, & \forall x \in \X \cup \mathit{Var}:\ \sigma'^-(x) = \sigma^-(x)\\
4.\, & \forall i:\ \sigma'^-(\pi_i) =\ \sigma^-(\pi_i)\\
\end{array} \right.
\]
{\sf Timeout Increment Transition: } For the edges $\nu_0^i$, if there is an edge in the TTD for process $P_i$ connecting source location $l_j^i$ to target location $l_k^i$ and is labeled by the instruction $\rho \Rightarrow \lan update^-_i, \gamma, f \ran$, then
\[
\nu_0^i \equiv (\sigma^-, \sigma'^-) \in  \Gamma^-_0 \Leftrightarrow \left\{
\begin{array}{ll} 1.\, &
\rho \mbox{ holds in } \sigma^- \\
2.\, &  \mathbf{If}\, \sigma^-(\tau_i) = 0\, \mathbf{then}\\
& \ \ \ \ \sigma'^-(\tau_i) = update^-_i > 0\, \\
&\mathbf{ else}\, \sigma'^-(\tau_i) = \sigma^-(\tau_i)\\
3.\, & \forall y \in \gamma:\ \sigma'^-(y) = \sigma'^-(\tau_i) + \sigma^-(y)\mbox{ and}\\
& \forall x \in \X\setminus\gamma:\ \sigma'^-(x) = \sigma^-(x) \\
4.\, & \forall v \in G \cup L_i: \sigma'^-(v) = f(\sigma^-(v))\ \mbox{and}\\
& \forall v \in \mathit{Var}\setminus(G \cup L_i):\ \sigma'^-(v) =
\sigma^-(v)\\
5.\, & \sigma^-(\pi_i) =\ l_j^i \mbox{ and }
\sigma'^-(\pi_i) = l_k^i
\end{array}
\right.
\]
Observe that $update^-_i$ is a slight modification of $update_i$. If $update^-_i= k_1$ s.t. $l - z \prec k_1 \prec m - z'$, then $update^-_i$ arbitrarily selects a value $\delta$ such that $l - \sigma^-(z) \prec \delta \prec m - \sigma^-(z')$. If $update^-_i= k_2$ s.t. $k_2 \succ l - z$, then $update^-_i$ arbitrarily selects a value $\delta$ such that $\delta \succ l - \sigma^-(z)$, else if $update^-_i= \infty$, then it selects the largest possible constant defined by the system and returns $\delta$. If $update^-_i= \max(\cM)$, $update^-_i$ nondeterministically selects any integer $\delta$ in $[0, M+1]$, where $M$ is the maximum of all the integers in $\cM$. Unlike the local timing variables appearing in $\gamma$ in a (clocked) ToM, these timing variables incrementally capture the value of next timeout in a  clockless ToM. An observant reader can see that  the relative delay captured by these local timing variables between events are same in both those models.\\\\
{\sf Synchronous Communication} For a pair of processes $P_s, P_r$ having edges $(e_s, e_r)$ :
\[
\nu_{syn\_comm}^{sr} \equiv (\sigma^-, \sigma'^-) \in \Gamma^-_{syn\_comm} \Leftrightarrow \left\{
\begin{array}{ll} 1.\, &
\rho \mbox{ holds in } \sigma^-\\
2.\, &  \sigma'^-(\tau_s) = update^-_s > \sigma^-(\tau_s)\\
&  \sigma'^-(\tau_r) = update^-_r > \sigma^-(\tau_r)\\
3.\, & \forall y \in (\gamma):\ \sigma'^-(y) = \sigma'^-(\tau_s) + \sigma^-(y), \ \mbox{ and}\\
& \forall y' \in (\gamma'):\ \sigma'^-(y') = \sigma'^-(\tau_r) + \sigma^-(y')\ \mbox{ and}\\
& \forall x \in \X\setminus(\gamma \cup \gamma'):\ \sigma'^-(x) = \sigma^-(x) \\
4.\, & \sigma'^-(\bar{m}) = \sigma^-(m)\\
5.\, & \forall v \in G \cup L_s: \sigma'^-(v) = g(\sigma^-(v)),\ \mbox{ and}\\
& \forall v \in G \cup L_r: \sigma'^-(v) = h(\sigma^-(v))\ \mbox{ and}\\
& \forall v \in \mathit{Var}\setminus(G \cup L_r \cup L_s):\ \sigma'^-(v) =
\sigma^-(v)\\
6.\, & \sigma^-(\pi_s) =\ l_j^s, \sigma^-(\pi_r) =\ l_j^r \mbox{ and }\\
&  \sigma'^-(\pi_s) = l_k^s, \sigma'^-(\pi_r) =\ l_k^r
\end{array}
\right.
\]

\subsection{Calendar based Models: Clockless Modeling} \label{calclockless}

\subsubsection{Clockless Syntax} Similar to the ToM, calendar based models
can also be defined in a clockless manner. However we restrict the set $\U$ to, $$\U = \T \cup \X \cup \mathit{Var} \cup
\C,$$ where $update^-_i$ is defined using same rule as in the case of clockless ToM.

\subsubsection{Clockless Semantics} Similar to the clockless ToM, we can define a transition system for clockless calendar based models. Here we need to modify the Time Progress, Timeout Increment, Send, and Receive Transitions as defined earlier for CCTS. Synchronous Communication transition is similar to the one for timeout based model with clockless semantics.\\\\
{\sf Time Progress Transition: } The first kind of edges $\nu_+$ are redefined so that all the timeout and calendar delay entries are decremented by the minimum of all timeouts and the message
delays in calendar. In particular,
\[
\nu_+ \equiv (\sigma^-, \sigma'^-) \in \Gamma^-_+ \Leftrightarrow \left\{ \begin{array}{ll}
1.\, & \min\{\sigma^-(\T) \cup \Delta(\sigma^-(\C))\} > 0 \\
2.\, & \forall \tau \in \T:\  \sigma'^-(\tau) = \sigma^-(\tau) - \min\{\sigma^-(\T) \cup \Delta(\sigma^-(\C))\}\\
3.\, & \forall \lambda \in \Delta(\sigma^-(\C)):\  \sigma'^-(\lambda) = \sigma^-(\lambda) - \min\{\sigma^-(\T) \cup \Delta(\sigma^-(\C))\}\\
4.\, & \forall x \in \X \cup \mathit{Var}:\ \sigma'^-(x) = \sigma^-(x) \\
5.\, & \forall i:\ \sigma'^-(\pi_i) =\ \sigma^-(\pi_i)
\end{array} \right.
\]
{\sf Timeout Increment Transition:} For the second kind of edges
$\nu_0^i$, if there is an edge in process $P_i$ connecting source location $l_j^i$
to target location $l_k^i$ and is labeled by the instruction $\rho
\Rightarrow \lan \tau_i := update^-_i, \gamma, f \ran$, then
\[
\nu_0^i \equiv (\sigma^-, \sigma'^-) \in \Gamma^-_0 \Leftrightarrow \left\{
\begin{array}{ll} 1.\, &
\rho \mbox{ holds in } \sigma^- \\
2.\, &  \mathbf{If}\, \min\{\sigma^-(\T)\} = \sigma^-(\tau_i) = 0\\
& \ \ \ \mathbf{then }\ \sigma'^-(\tau_i) = update^-_i > 0\\
& \mathbf{else }\,\  \sigma'^-(\tau_i) = \sigma^-(\tau_i)\\
3.\, & \forall y \in \gamma:\ \sigma'^-(y) = \sigma'^-(\tau_i) + \sigma^-(y)\mbox{ and}\\
& \forall x \in \X\setminus\gamma:\ \sigma'^-(x) = \sigma^-(x)\\
4.\, & \forall v \in G \cup L_i: \sigma'^-(v) = f(\sigma^-(v))\ \mbox{and}\\
& \forall v \in \mathit{Var}\setminus(G \cup L_i):\ \sigma'^-(v) =
\sigma^-(v)\\
5.\, & \sigma^-(\pi_i) =\ l_j^i \mbox{ and } \sigma'^-(\pi_i) = l_k^i
\end{array}
\right.
\]
{\sf Send Transition:} If there is an edge in process $P_i$, which
connects source location $l_j^i$ to target location $l_k^i$ and is
labeled by the instruction $\rho \Rightarrow \lan send(m, i, \Omega,
\Lambda), update^-_i, \gamma, f \ran$, then we have corresponding
edge $\nu_{send}^i$ which adds $|\Omega|$ cells to the calendar array
$\C$:
\[
\nu_{send}^i \equiv (\sigma^-, \sigma'^-) \Leftrightarrow \left\{
\begin{array}{ll} 1.\, &
\rho \mbox{ holds in } \sigma^- \\
2.\, & \mathbf{If}\, \min\{\sigma^-(\T)\} = \sigma^-(\tau_i) = 0\\
& \ \ \ \mathbf{then }\ \sigma'^-(\tau_i) = update^-_i > 0\\
& \mathbf{else }\,\  \sigma'^-(\tau_i) = \sigma^-(\tau_i)\\
4.\, & \forall y \in \gamma:\ \sigma'^-(y) = \sigma'^-(\tau_i) + \sigma^-(y)\mbox{ and}\\
& \forall x \in \X\setminus\gamma:\ \sigma'^-(x) = \sigma^-(x)\\
5.\, & \forall v \in G \cup L_i: \sigma'^-(v) = f(\sigma^-(v))\ \mbox{and}\\
& \forall v \in \mathit{Var}\setminus(G \cup L_i):\ \sigma'^-(v) = \sigma^-(v)\\
6.\, & \forall (r, \lambda_r) \in \Omega: \sigma'^-(\C) := \sigma^-(\C) + \mbox{{\sf \{m, i, r, $\lambda_r$\}}}\\
7.\, & \sigma^-(\pi_i) =\ l_j^i \mbox{ and } \sigma'^-(\pi_i) = l_k^i
\end{array}
\right.
\]
{\sf Receive Transition:} If there is an edge in process
$P_r$, which connects source location $l_j^r$ to target location
$l_k^r$ and is labeled by the instruction $True \Rightarrow \lan
receive(m, i, r), \gamma, f \ran$, then we have corresponding edge
$\nu_{receive}^r$ which deletes the cell containing $\{m, i, r, \lambda_r\}$
from the calendar array $\C$:
\[
\nu_{receive}^r \equiv (\sigma^-, \sigma'^-) \Leftrightarrow \left\{
\begin{array}{ll} 1.\, & \exists \{m, i, r, \lambda_r\} \in \sigma^-(\C)\ \mbox{s.t. } \lambda_r = 0\\
2.\, &  \sigma'^-(\tau_r) = update^-_r > 0\\
3.\, & \forall y \in \gamma:\ \sigma'^-(y) = \sigma'^-(\tau_i) + \sigma^-(y)\ \mbox{and}\\
& \forall x \in \X\setminus\gamma:\ \sigma'^-(x) = \sigma^-(x)\\
4.\, & \forall v \in G \cup L_r: \sigma'^-(v) = f(\sigma^-(v))\ \mbox{and}\\
& \forall v \in \mathit{Var}\setminus(G \cup L_r):\ \sigma'^-(v) = \sigma^-(v)\\
5.\, & \sigma'^-(\C) := \sigma^-(\C)\setminus\{m, i, r, \lambda_r\}\\
6.\, & \sigma^-(\pi_r) =\ l_j^r \mbox{ and } \sigma'^-(\pi_r) = l_k^r
\end{array}
\right.
\]
Thus the clockless semantics defines a possible {\em clockless computation} $\xi^-$ of TCTS/CCTS as a
sequence of states $\sigma_0^-, \sigma^-_1, \cdots$.

\subsection{LTL formulas for Clockless Models} \label{LTL-classic-semantics}
A remark about the LTL formulas that would be verified against clockless models, is in order. These formulas will not involve the global timing variable $t$. The  LTL formulas will be built using finitely many atomic propositions (constraints), which may be defined in terms of state variables for which the possible combinations of valuations needs to be finite.

Assuming that typical arithmetic constraints are defined in terms of variables in $\U$ (as defined before for clockless timeout and calender models), let us now define a point-wise or event based semantics for LTL formulas based on its classical semantics~\cite{GGP99}.  A model for a LTL formula would consist of a sequence of states of the form $$ \sigma_0, \sigma_1, \cdots, $$
such that each state $\sigma_i$ gives a boolean interpretation ($\true, \false)$ to the propositions, and non-negative integer valued interpretation to the timeout variables in $\T$, timing variables in $\X$, and state variables in $\mathit{Var}$, all of which are  bounded above by some positive integer constant. In a state $\sigma_i$, let us assume $\sigma_i(v)$ to be the value of $v \in \U$. Considering an example of an arithmetic constraint as $t_j - t_k \geq c$, where $t_j, t_k \in T \cup \X$ and $c$ an integer constant, the satisfaction relation $\models$ can be defined as
\[
\begin{array}{ccc}
\sigma_i \models p & {\rm iff} &  \sigma_i(p) = \true \\
\sigma_i \models t_j - t_k \geq c & {\rm iff} & {\sigma}_i(t_j) - {\sigma}_i(t_k) \geq c \\
\sigma_i \models \neg \phi & {\rm iff} & \sigma_i \not \models \phi \\
\sigma_i \models \phi \vee \psi & {\rm iff} & \sigma_i \models \phi \,\, {\rm or} \,\, \sigma_i \models \psi \\
\sigma_i \models \bigcirc \phi & {\rm iff} & \sigma_{i+1} \models \phi \\
\sigma_i \models \phi \U \psi & {\rm iff} & \exists k > i.\, \sigma_k \models \psi \,\, {\rm and} \,\, \forall j. i \leq j < k.\, \sigma_j \models \psi
\end{array}
\]
In terms of these LTL formulas, using Clockless ToM, one can essentially verify all those qualitative properties of the associated real-time system, which are otherwise prohibitively difficult to do using the clocked ToM models and timed temporal logics. This is because clockless models preserve the qualitative behavior of the clocked models and LTL can effectively specify these properties. As the valuations of the variables in the clockless models are bounded, the clockless models effectively give rise to finite state behaviors. Indeed, we can also estimate the approximate size of the clockless TCTS having direct bearing on the time complexity of its LTL model-checking. Assume a clockless ToM with $n$ parallel processes with $k$ local timing variables. Let the valuations of timeouts and timing variables be bounded above by $M = \max(\cM)$. Also let the sizes of the clockless TTDs of these processes are bounded by $D$. In terms of these, the size of the clockless TTS could be bounded by ${\cal F} = {\mathbf O}(\max\{M^{n+k}D^n, |\Gamma^-|\})$, using asymptotic notation. This, in turn implies that complexity of model checking such clockless TTS for a LTL formula $\phi$ would be ${\mathbf O}({\cal F}2^{|\phi|})$~\cite{VW86}.


\subsection{Clockless Models (Bi-)Simulate Clock Models}
In this section we will show that clockless models (bi-)simulate clock models with respect to LTL formulas. Let us consider a ToM $P$ and its TCTS $S_P = (\V, \Sigma, \Sigma_0, \Gamma)$ and also the clockless ToM $P^-$ and corresponding timeout based clockless transition system $S^-_{P} = (\V^-, \Sigma^-, \Sigma^-_0, \Gamma^-)$; both of them modeling the same system. 
Given a computation $\xi : \sigma_0 \ra \sigma_1 \ra \cdots$ over $S_{P}$ 
let us generate a clockless computation as a sequence of states $ \sigma^-_0, \sigma^-_1 \cdots$ over $S^-_{P}$ as follows:
\begin{itemize}
\item Initial states correspond: 
%
\[
\begin{array}{l}
\forall \tau \in \T.\ \sigma^-_0(\tau) = \sigma_0(\tau),\\
\forall x \in \X.\ \sigma^-_0(x) =  \sigma_0(x), \\
\sigma^-_0(\pi_i) = \sigma_0(\pi_i) =  \perp.
\end{array}
\]
    \item Entry transition: if $(\sigma_0, \sigma_1) \in \Gamma_e$ then
    \[ \left\{
    \begin{array}{ll}
    1. & \forall \tau \in \T. \sigma^-_1(\tau) = \sigma_1(\tau),    \\
    2. & \forall x \in \X. \sigma^-_1(x) = \sigma_1(x),\\
    3. & \sigma^-_1(\pi_i) = \sigma_1(\pi_i) = l^i_0
    \end{array}
    \right. \]
    \item Time progress transition: if $(\sigma_{i-1}, \sigma_i) \in \Gamma_+$ then
    \[ \left\{
    \begin{array}{ll}
    1. & \forall \tau \in \T. \sigma^-_i(\tau) = \sigma_i(\tau) - \min \{\sigma_{i-1}(\T)\},    \\
    2. & \forall x \in \X. \sigma^-_i(x) = \sigma_i(x),\\
    3. & \forall i.\sigma^-_i(\pi_i) = \sigma_i(\pi_i).
    \end{array}
    \right. \]
    \item Timeout increment transition: if $(\sigma_{i-1}, \sigma_i) \in \Gamma_e$ (which is labeled by the instruction $\rho
\Rightarrow \lan \tau_i := update^-_i, \gamma, f \ran$) then
    \[ \left\{
    \begin{array}{ll}
    1. & \,\, \mbox{if}\,\, \sigma_{i-1}^-(\tau_i) = 0\,\, \mbox{then} \,\, \sigma^-_i(\tau_i) = {update}^{-}_i, \,\,\mbox{else} \,\, \sigma_i^-(\tau_j) = \sigma_i(\tau_j)     \\
    2. & \forall x \in \gamma. \sigma^-_i(x) = \min \{ \sigma_{i-2}(\T) \} + \sigma_{i-1}(x),\\
    3. & \forall x \in \X \setminus \gamma. \sigma^-_i(x) =  \sigma_{i-1}(x),\\
    4. & \forall i.\sigma^-_i(\pi_i) = \sigma_i(\pi_i).
    \end{array}
    \right. \]
    where $update^{-}_i$ is defined in $P^-$.
\item Synchronous communication: if $(\sigma_{i-1}, \sigma_i) \in \Gamma_{syn\_comm}$ then
\[ \left\{
\begin{array}{ll}
1. &  \sigma^-_i(\tau_s) = \sigma_i(\tau_s) \,\, \mbox {and}\   \sigma^-_i(\tau_r) = \sigma_i(\tau_r)\\
2.\, & \forall x \in \X. \sigma^-_i(x) = \sigma_i(x) \\
3.\, & \sigma^-_i(\bar{m}) = \sigma_i(\bar{m}) \,\, \mbox{and}\,\, \sigma^-_{i-1}(m) = \sigma_{i-1}(m)\\
4.\, & \forall v \in G \cup L_s: \sigma^-_i(v) = \sigma_i(v)\ \mbox{and}\ \forall v \in G \cup L_r: \sigma^-_i(v) = \sigma_i(v)\\
& \forall v \in \mathit{Var}\setminus(G \cup L_s \cup L_r):\ \sigma^-_i(v) = \sigma_i(v)\\
5.\, & \sigma^-_{i-1}(\pi_s) =\ \sigma_{i-1}(\pi_s) =\ l_j^s, \sigma^-_{i-1}(\pi_r) =\ \sigma_{i-1}(\pi_r) =\ l_j^r \mbox{ and }\\
&  \sigma^-_i(\pi_s) =\ \sigma^-_i(\pi_s) =\ l_k^s, \sigma^-_i(\pi_r) =\ \sigma_i(\pi_r) =\ l_k^r
\end{array}
\right.
\]
\end{itemize}
Check that $\sigma^-_0 \in \Sigma^-_0$ and $\forall i. (\sigma^-_{i-1}, \sigma^-_i) \in \Gamma^-$. It is clear $\xi^- = \sigma^-_0 \ra \sigma^-_1 \ra \ldots$ forms a clockless computation over $S^-_{P}$. We can associate a mapping $\Tr: \Sigma \times \Sigma \rightarrow \Sigma^-$ parameterized by an entry transition as follows. Fix two states, $\sigma_0 \in \Sigma_0, \sigma_1 \in \Sigma$, such that $(\sigma_0, \sigma_1) \in \Gamma_e$. Call $\gamma = (\sigma_0, \sigma_1)$. Then define  ${\Tr}_{\gamma}(\sigma_0, \sigma_0) = \sigma^-_0, {\Tr}_{\gamma}(\sigma_{i}, \sigma_{i-1}) = \sigma_i^-,\, \forall i \geq 1$.

We say that computations $\xi: \sigma_0\sigma_1 \ldots$ in $S_P$ and $\xi^-: \sigma^-_0\sigma^-_1 \ldots$ in $S^-_{P}$ correspond if and only if there exists $\Tr_{\gamma}: \Sigma \times \Sigma \rightarrow \Sigma^-$ such that $\sigma^-_0 = {\Tr}_{\gamma}(\sigma_0, \sigma_0)$ and for every $i \geq 0,\, \sigma^-_i = {\Tr}_{\gamma}(\sigma_i, \sigma_{i-1})$, where $\gamma = (\sigma_0, \sigma_1)$. Let $\sigma \in \Sigma$ and $\sigma^- \in \Sigma^-$ be two states and there be a computation in $S_P$ which starts in $\sigma$. Then it is easy to see that there exists a corresponding computation in $S^-_{P}$ beginning with $\sigma^-$~\cite{GGP99}.

We consider LTL formulas consisting of propositions and  variables appearing in clockless transition system of $S^-_{P}$. Assume $\sigma \in \Sigma$ and $\sigma^- \in \Sigma^-$ are two states such that ${\Tr}_{\gamma}(\sigma, \sigma') = \sigma^-$ for some $\sigma' \in \Sigma$ and some entry transition $\gamma$. Then for any LTL formula $\phi$, $\sigma^- \models \phi$ implies $\sigma \models \phi$ (using the semantics of LTL formulas as discussed in Section~\ref{LTL-classic-semantics}). This can be proved using the induction on the structure of $\phi$. Finally, $S^-_{P} \models \phi$ implies $S_{P} \models \phi$. This is in some sense, we can say $S^-_{P}$ simulates $S_{P}$~\cite{GGP99}. Thus it is enough to verify properties on the clockless transition system $S^-_{P}$ instead of on $S_{P}$.

Similar results can be established for calendar-based clocked transition system (CCTS) also.
In fact a reverse mapping cane be defined too. To see this let us assume $\xi^- = \sigma^-_0, \sigma^-_1 \ldots$ to be a clockless computation over $S^-$. Now generate a sequence of states $\sigma_0, \sigma_1 \ldots$ as follows.
\begin{itemize}
\item $\sigma_0(t) = \min\{ \sigma^-_0(\T)\}, \forall \tau \in \T. \sigma_0(\tau) = \sigma^-_0(\tau), \forall x \in \X. \sigma_0(x) = \sigma^-_0(x), \sigma_0(\pi_i) = \sigma^-_0(\pi_i) = \perp$.
    \item \[ \mbox{if} \,\, (\sigma^-_0, \sigma^-_1) \in \Gamma_e \,\, \mbox{then} \,\, \left\{
    \begin{array}{ll}
    1. & \forall \tau \in \T. \sigma_1(\tau) = \sigma^-_1(\tau),    \\
    2. & \forall x \in \X. \sigma_1(x) = \sigma^-_1(x),\\
    3. & \sigma_1(\pi_i) = \sigma^-_1(\pi_i) = l^i_0,\\
    4. & \sigma_1(t) = \sigma^-_1(t) = \sigma^-_0(t)
    \end{array}
    \right. \]
    \item \[ \mbox{if} \,\, (\sigma^-_{i-1}, \sigma^-_i) \in \Gamma_+ \,\, \mbox{then} \,\, \left\{
    \begin{array}{ll}
    1. & \forall \tau \in \T. \sigma_i(\tau) = \sigma_{i-1}(\tau),    \\
    2. & \forall x \in \X. \sigma_i(x) = \sigma_{i-1}(x),\\
    3. & \forall i.\sigma_i(\pi_i) = \sigma^-_i(\pi_i),\\
    4. & \sigma_i(t) = \min \{\sigma_{i-1}(\T) \}.
    \end{array}
    \right. \]
    \item \[ \mbox{if} \,\, (\sigma^-_{i-1}, \sigma^-_i) \in \Gamma_e \,\, \mbox{then} \,\, \left\{
    \begin{array}{ll}
    1. & \,\, \mbox{if}\,\, \sigma_{i}^-(\tau_i) = 0\,\, \mbox{then} \,\, \sigma_i(\tau_i) = {update}_i, \,\,\mbox{else} \,\, \sigma_i(\tau_j) = \sigma_{i-1}(\tau_j)     \\
    2. & \forall x \in \X. \sigma_i(x) =  \sigma_{i-1}(x),\\
    3. & \forall i.\sigma_i(\pi_i) = \sigma^-_i(\pi_i),\\
    4. & \sigma_i(t) = \sigma_{i-1}(t)
    \end{array}
    \right. \]
\item \[ \mbox{if} \,\, (\sigma^-_{i-1}, \sigma^-_i) \in \Gamma_{syn\_comm} \,\, \mbox{then} \,\, \left\{
\begin{array}{ll}
1. &  \sigma_i(\tau_s) = \sigma^-_i(\tau_s) \,\, \mbox {and}\   \sigma_i(\tau_r) = \sigma^-_i(\tau_r)\\
2.\, & \forall x \in \X. \sigma_i(x) = \sigma^-_i(x) \\
3.\, & \sigma_i(\bar{m}) = \sigma^-_i(\bar{m}) \,\, \mbox{and}\,\, \sigma_{i-1}(m) = \sigma^-_{i-1}(m)\\
4.\, & \forall v \in G \cup L_s: \sigma_i(v) = \sigma^-_i(v)\ \mbox{and}\ \forall v \in G \cup L_r: \sigma_i(v) = \sigma^-_i(v)\\
& \forall v \in \mathit{Var}\setminus(G \cup L_s \cup L_r):\ \sigma_i(v) = \sigma^-_i(v)\\
5.\, & \sigma_{i-1}(\pi_s) =\ \sigma^-_{i-1}(\pi_s) =\ l_j^s, \sigma_{i-1}(\pi_r) =\ \sigma^-_{i-1}(\pi_r) =\ l_j^r \mbox{ and }\\
&  \sigma^-_i(\pi_s) =\ \sigma^-_i(\pi_s) =\ l_k^s, \sigma^-_i(\pi_r) =\ \sigma_i(\pi_r) =\ l_k^r,\\
6. & \sigma_i(t) = \sigma_{i-1}(t).
\end{array}
\right.
\]
\end{itemize}
Clearly, $\xi: \sigma_0 \ra \sigma_1 \ra \cdots$ is a computation over $S$. Associate a mapping $\Tr': \Sigma^-
\rightarrow \Sigma$ with this such that $\Tr': \sigma^-_i \mapsto \sigma_i, \forall i$.
Let us try to compose these two mappings. Note that $\Tr \circ \Tr' = {\rm id}, \Tr' \circ \Tr = {\rm id}$ where ${\rm id}$  is an identity mapping. This implies that $\Tr$ is a bijective mapping and $(\Tr)^{-1} = \Tr'$.

Define a relation $\cB \subseteq \Sigma \times \Sigma^-$ as follows: for two states $s \in \Sigma$ and $s^- \in \Sigma^-$ we have $\cB(s,s^-)$ if and only if $s^- = \Tr(s)$. Assume $s$ and $s^-$ satisfy the same atomic propositions. Also observe that
\begin{itemize}
\item for every state $s_1 \in \Sigma: (s,s_1) \in \Gamma$ there exists $s^-_1 \in \Sigma^-: (s^-, s^-_1) \in \Gamma^-$  such that $s^-_1 = \Tr(s_1)$, {\em i.e.,} $\cB(s^-,s^-_1)$.
    \item for every state $s^-_1 \in \Sigma^-:(s^-,s^-_1) \in \Gamma^-$ there exists $s_1 \in \Sigma: (s, s_1) \in \Gamma$  such that $s_1 = (\Tr)^{-1}(s^-_1)$, {\em i.e.,} $\cB(s^-,s^-_1)$.
        \end{itemize}
        Hence $\cB$ is a bisimulation relation between $S$ and $S^-$.
Finally, we can see for this bisimulation relation $\cB$, for every initial state $s_0 \in \Sigma$ in $S$ there
is an initial state $s^-_0 \in \Sigma^-$ in $S^-$ such that $cB(s_0,s^-_0)$. In addition, for every initial state
$s^-_0 \in \Sigma^-$ in $S^-$ there is an initial state $s_0 \in \Sigma$ in $S$ such that $\cB(s_0,s^-_0)$. Hence
$S$ and $S^-$ are bisimulation equivalent~\cite{GGP99}. Since bisimulation equivalent structures preserve LTL formulas~\cite{GGP99} we shall be dealing with clockless timeout based models for our verification purposes.

\section{Experimental Evaluation}\label{results}

In this section we illustrate finite state verification of real-time systems through clockless modeling on three real-time protocols introduced earlier - Fisher's Mutual Exclusion Protocol, TGC, and TTA startup protocol.
We perform finite state model checking of these protocols by {\sf Spin} and {\sf SAL-smc} model checkers.
For applying our technique we assume that the timeout increments of these protocols are more than one time unit.
We carry out our experiments on a machine with 2.26GHz Intel Core 2 Duo processor, 3 MB shared level 2 cache and 2GB 1066MHz DDR3 SDRAM, running MAC OS X Version 10.5.7.
For experimentation with {\sf Spin}, we use {\sf XSpin} graphical interface.
To verify a property {\sf prop} for a {\sf SAL} specification \textit{model.sal} we use the following {\sf SAL} command:
\begin{center}
{\sf sal-smc -v 3 model prop --enable-dynamic-reorder}\\
\end{center}
Here {\sf enable-dynamic-reorder} is a flag used with {\sf SAL-smc} that enables dynamic reordering of BDD variables.

\subsection{Fischer's Mutual Exclusion Protocol}
\begin{figure}
\centering
\includegraphics[width=5.0in]{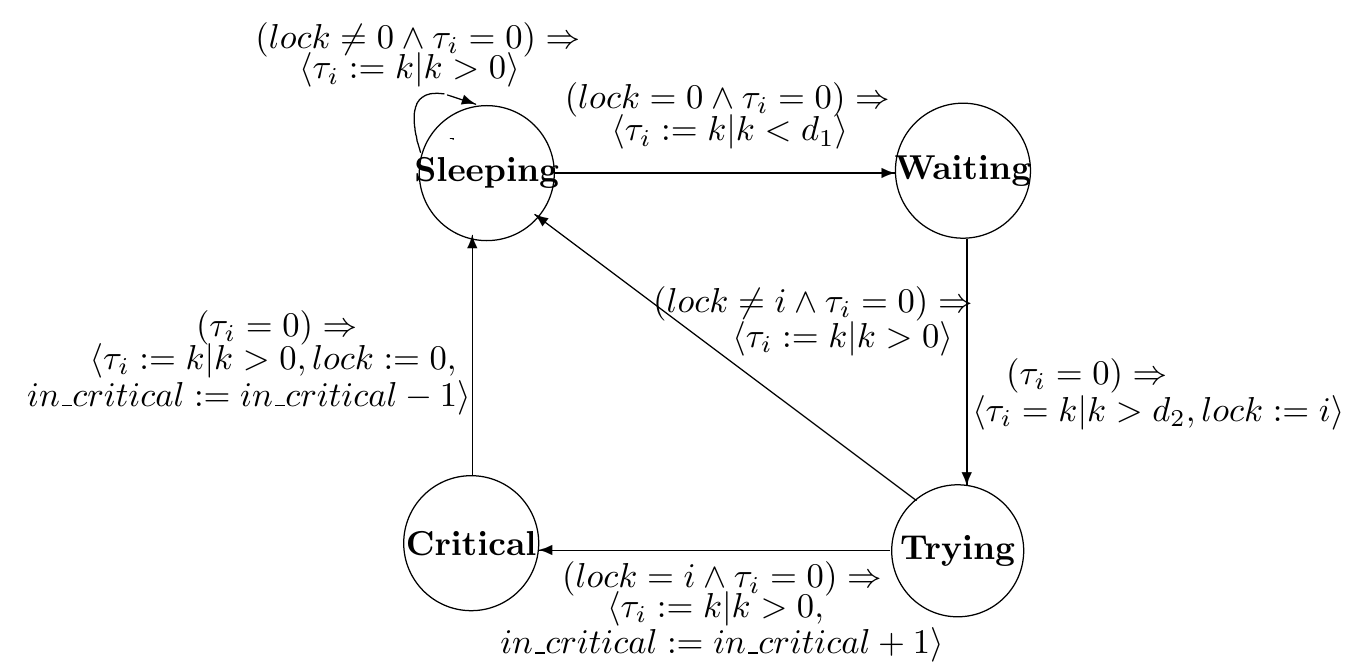}
\caption{Clockless model for the $i^{th}$ processor in the Fischer's Protocol} \label{fig:fig6}
\end{figure}

A clockless model of the Fisher's mutual exclusion protocol is depicted in Figure~\ref{fig:fig6}. We consider the following {\tt safety} property for Fischer's protocol,  {\textit {``no more
than one processor can be in the critical region at any time''}}.  The property is frequently referred as {\tt mutual exclusion} property. This can be represented in LTL as: $$\Box (in\_critical \ \leq 1)$$

To verify the {\tt safety} property for Fischer's mutual exclusion
protocol in {\sf Spin} we used \emph{exhaustive verification} and \emph{bitstate hashing}
technique available in {\sf Spin}, in both the cases keeping the the option
of \emph{partial order reduction} turned on. By \emph{exhaustive verification}
technique, we could verify models containing only upto
$4$ nodes. \emph{Bitstate hashing} enabled us to verify the same property for
models with upto $6$ nodes. Table~\ref{tab4} illustrates the
computational resources and time required to prove the {\tt safety}
property for Fischer's mutual exclusion protocol using \emph{bitstate
hashing} technique.
\begin{table}
\caption{Computational resources required for verification of the Fischer's Protocol using \emph{bitstate hashing}} \label{tab4}
\begin{center}
\begin{tabular*}{0.7\textwidth}{@{\extracolsep{\fill}}||c|c|c|c|c|c|c||}
\hline \hline Property & N & \# States & \# States & \# Transitions & Memory & Time \\
&  & Stored & Matched &  & (MB) & (sec) \\
\hline
\hline & 2 & 563 & 463 & 1026 & 8.501 & 0.1\\
\cline{2-7} & 3 & 18220 & 29625 & 47845 & 8.598 & 0.11\\
\cline{2-7}  {\tt Safety} & 4 & 667995 & 1716011& 2384003 & 44.383 & 5.01 \\
\cline{2-7} & 5 & 21373206 & 75073507 & 96446713 & 395.366 & 203.09\\
\cline{2-7} & 6 & 36720364  & 1.4129329e+08 & 1.7801365e+08 & 1722.014 & 908.56\\
\hline \hline
\end{tabular*}
\end{center}
\end{table}


\begin{table}
\caption{States explored and time required to verify mutual-exclusion property by {\sf SAL-smc} for Fischer's protocol} \label{tab1}
\begin{center}
{\footnotesize
\begin{tabular*}{0.58\textwidth}{@{\extracolsep{\fill}}||c|c|c||c|c|c||}
\hline \hline \# Nodes & \# States & Time & \# Nodes & \# States & Time\\
 & Explored & (sec) & & Explored & (sec) \\
\hline
2 & 468 & 0.15& 10 & 1.189e12 & 69.9\\ \hline
3 & 7968 & 0.30& 11 & 1.697e13& 213.76\\ \hline
4 & 124760& 1.40 & 12 & 2.417e14 & 196.36\\ \hline
5 & 1.876e6 & 2.35 & 13 & 3.438e15& 2767.91 \\ \hline
6 & 2.760e7 & 3.84 & 14 & 4.885e16& 21731.91\\ \hline
7 & 4.010e8 & 12.43 & 15 & 6.935e17 & 4516.85 \\ \hline
8 & 5.786e9 & 23.04 & 16 & 9.839e18 & 10376.53\\ \hline
9 & 8.306e10 & 44.604 & 17 & -- & --\\
\hline \hline
\end{tabular*}}
\end{center}
\end{table}

We perform clockless modeling of Fischer's protocol in {\sf SAL} language. Table~\ref{tab1} presents the number of states visited and time required to prove the {\tt mutual exclusion} property. We have been able to verify the {\tt mutual exclusion} property for Fischer's protocol with $16$ processors in around $3$ hours (except the model for 14 nodes, which took around 6 hours). We tried to verify the protocol for 17 and 18 nodes, and in both the cases, verification ran for more than 7 hours. We did not go for  higher number of nodes.

The Fisher's protocol has been verified under dense time for the same {\tt mutual exclusion}
property in~\cite{DS04a}. A direct attempt to prove the property by $k$-induction with induction depth up to $15$ fails for even $2$ processors.
However, using a sequence of lemmas it was possible to prove the property by induction at depth $1$ for upto $13$ processors for the same {\sf SAL} specification (Table 3.1 of~\cite{DS04a}). The property was also proved by induction by a sequence of lemmas for a  \textit{different} {\sf SAL} specification for a maximum number of $53$ processors (Table 3.5 of~\cite{DS04a}).

To compare the performance and scalability of our verification approach with {\sf UPPAAL}, we verified Fischer's mutual exclusion protocol available with {\sf UPPAAL} distribution. The UPPAAL model is based on the framework of  \emph{timed automaton}. The {\tt mutual exclusion} property could be verified successfully for up to $12$ nodes. For $13$ nodes, the verification process did not stop even in $7$ hours. In verification with UPPAAL, the TA is reduced to the zone automata which are finite representations of infinite state systems. Although both our clockless verification scheme and {\sf UPPAAL's} zone automata based verification are based on abstracting an infinite system to a finite one, this experimental result shows that our technique is more scalable than {\sf UPPAAL}, while using {\sf SAL-smc} model checker.

\subsection{Train-Gate Controller}
\begin{figure}
\centering
\includegraphics[scale=0.8]{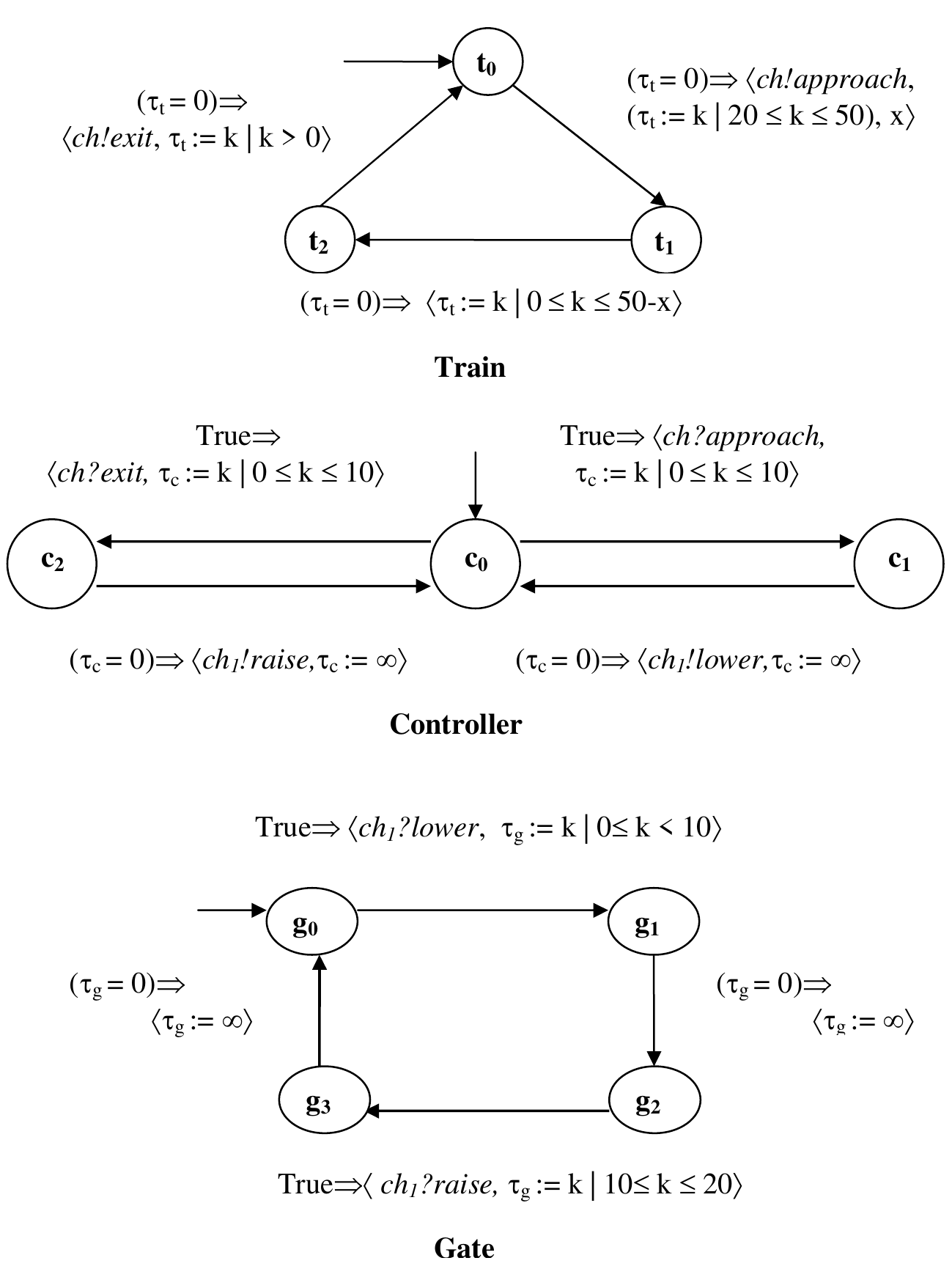}
\caption{Clockless model for Train-Gate Controller}
\label{fig:fig4}
\end{figure}
A clockless model of TGC is depicted in Figures~\ref{fig:fig4}.
For the TGC example as discussed before, we consider {\tt safety} and {\tt timeliness} properties for verification. The {\tt safety}
property says:  \textit {When the Train crosses the line, the Gate should be down}. The property is expressed in LTL as: $$\Box ((t\_state = t_2) \imply (g\_state = g_2))$$ where, $t\_state$ denotes different states of the Train, and it is $t_2$, when it comes into the crossing, $g\_state$ denotes different states of Gate, and is $g_2$, when the Gate is down.

{\tt Timeliness} property, in general ensures that the time between two states will by bounded by a particular value. We can find many
{\tt timeliness} properties in this example. We select an important one, \textit{``the time between the transmission of the $approach$ signal by
the Train and when the Gate is down should not be more than $20$ time units''}.
To verify this property we use two auxiliary flags, $flag_1$ and $flag_2$ in
our model. When the first event occurs $flag_1$ is set as $true$.
When the second event happens, $flag_2$ is set as $true$ and
$flag_1$ is reset to $false$.

A global variable $time\_\mathit{diff}$ initially set to $0$, captures the time between the instants when two flags are set.
During every discrete transition between the two discrete transitions of interest, minimum timeout value is added to
$time\_\mathit{diff}$. The {\tt timeliness} property is then specified as follows,  \textit{``the value of   $time\_\mathit{diff}$ never goes above 20''}. This is expressed in LTL as, $$\Box (time\_\mathit{diff} \ \leq \ 20)$$
In Table~\ref{tab5}, we illustrates computational resources and time required to prove the {\tt safety} and the {\tt timeliness} properties for TGC by {\sf Spin} model checker. Both the properties have been proved by \emph{exhaustive verification} keeping the the option of \emph{partial order reduction} turned on.

\begin{table} 
 \caption{Computational resources and time required for verification of the Train-Gate Controller under exhaustive verification} \label{tab5}
\begin{center}
\begin{tabular*}{0.6\textwidth}{@{\extracolsep{\fill}}||c|c|c|c|c|c||}
\hline \hline Properties & \# States & \# States & \# Transitions & Memory & Time \\
 & Stored & Matched &  & (MB) & (sec) \\
\hline
\hline {\tt Safety} & 246236 & 422596 & 668832 & 47.947 & 1.50\\
\hline {\tt Timeliness} & 253500 & 415484 & 668984 & 50.389 & 1.58\\
\hline \hline
\end{tabular*}
\end{center}
\end{table}

We verify the {\tt safety} and {\tt timeliness} properties for TGC by {\sf SAL-smc}, and the result is shown in Table~\ref{tab2}.

\begin{table} 
 \caption{States explored and time required to verify {\tt safety} and {\tt timeliness} properties by {\sf SAL-smc} for TGC} \label{tab2}
\begin{center}
\begin{tabular*}{0.3\textwidth}{@{\extracolsep{\fill}}||c|c|c||}
\hline \hline Properties & \# States & Time \\
 & Explored & (sec)\\
\hline {\tt Safety} &  1.123e6 & 5.24\\
\hline {\tt Timeliness} & 4.807e5 & 2.41\\
\hline \hline
\end{tabular*}
\end{center}
\end{table}
It may be noted that dense time verification of the {\tt safety} property for TGC took 46.15 seconds~\cite{DS04a}. This was proved by k-induction at depth 14 using {\sf SAL-inf-bmc}.

\subsection{TTA Startup Algorithm}
\begin{figure}
\centering
\includegraphics[scale=0.7]{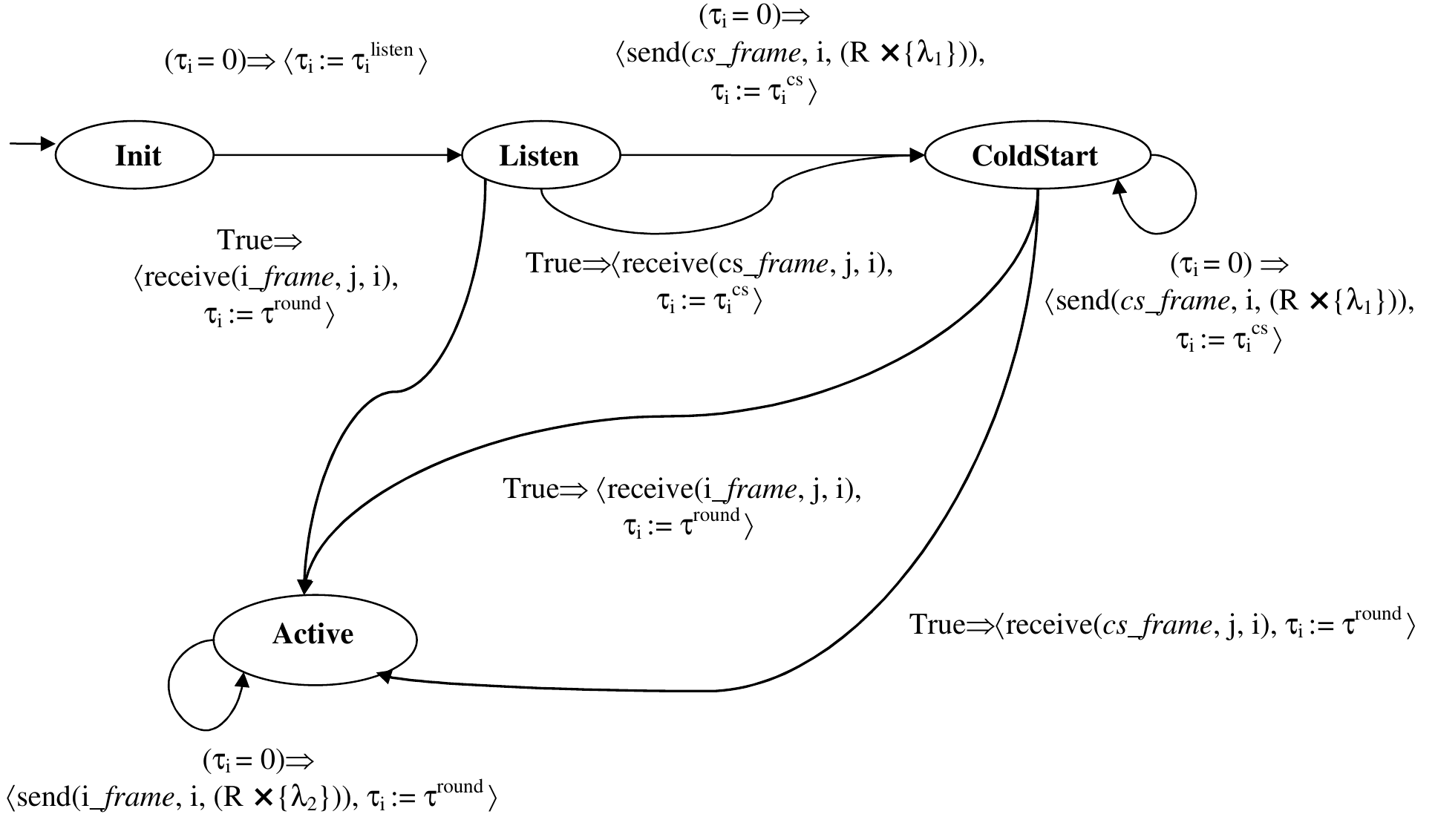}
\caption{Clockless model for the $i^{th}$ processor in TTA Startup
algorithm.} \label{fig:fig5}
\end{figure}
Figure~\ref{fig:fig5} depicts the clockless model for the TTA startup algorithm as discussed before in the Section~\ref{clocked-tta}.
We consider the following {\tt safety} property, \textit{``whenever any two nodes are in their \emph{active} state the nodes agree on the slot time''}.
For two nodes participating in the startup process, the corresponding LTL property is given below:
$$\Box ((p_1 \wedge p_2) \wedge (q_1 \wedge q_2)\ \imply\ \Diamond (r \wedge s)),$$
where $p_1 \equiv (pc[0] = state\_active), p_2 \equiv (pc[1] = state\_active)$, $q_1 \equiv (time\_out[0] > 0)$, $q_2 \equiv (time\_out[1] > 0)$, $r \equiv (time\_out[0] = time\_out[1])$, $s \equiv (slot[0] = slot[1])$. Also, $pc[i]$ denotes the current state of the $i^{th}$ node, $time\_out[i]$ denotes the timeout of the $i^{th}$ node, and $slot[i]$ denotes the current time slot viewed by the $i^{th}$ node. $state\_active$ characterizes the $synchronized$ state of a node.

The {\tt safety} property ensures that when the nodes are in $active$ state, then they are indeed synchronized. But it does not address the question whether all the nodes will be eventually synchronized or not.
To ensure that this happens, it is specified in the form of the following {\tt liveness} property,  \textit{``eventually all the nodes will be in $active$ state and continue to do so''}.
This {\tt liveness} property for two nodes can be specified in LTL as follows:
$$ \Diamond \Box ((pc[0] = state\_active) \wedge (pc[1] = state\_active))$$
To verify the {\tt safety} and the {\tt liveness} property for TTA startup in {\sf Spin}, we use
 both \emph{exhaustive verification} and \emph{bitstate hashing} techniques with \emph{partial
order reduction} availed. By \emph{exhaustive verification} technique, the {\tt safety} property
can be verified for TTA models containing upto $5$ nodes, and the
{\tt liveness} property can be verified upto $4$ nodes. \emph{Bitstate hashing}
enables us to verify both the properties for models with upto $9$
nodes. For $10$ nodes, the verification does not terminate even in
$4$ hours. Table~\ref{tab6} illustrates the computational
resources and time required to prove the {\tt safety} and {\tt liveness}
properties for TTA Startup protocol using \emph{bitstate hashing}
technique.


\begin{table}
 \caption{Computational resources and time required to verify {\tt safety} and {\tt liveness} property by bitstate hashing technique in Spin for TTA Startup} \label{tab6}
\begin{center}
\begin{tabular*}{0.72\textwidth}{@{\extracolsep{\fill}}||c|c|c|c|c|c|c||}
\hline \hline Properties & N & \# States & \# States & \# Transitions & Memory & Time \\
&  & Stored & Matched &  & (MB) & (sec)\\
\hline \hline        & 2 & 487 & 143 & 630 & 8.501 & 0.01\\
\cline{2-7}        & 3 & 6142 & 6490 & 12632 & 8.501 & 0.05\\
\cline{2-7}        & 4 & 217852 & 483497 & 701349 & 8.501 & 1.46\\
\cline{2-7}        & 5 & 4126813 & 13188075 & 17314888 & 8.501 & 34.72\\
\cline{2-7} {\tt Safety} & 6 & 16508262 & 62593403 & 79101665 & 8.501 & 165.46\\
\cline{2-7}        & 7 & 34442659 & 1.2702415e+08 & 1.6146681e+08 & 8.501 & 364.99\\
\cline{2-7}        & 8 & 40175448 & 2.4473144e+08 & 2.8490689e+08 & 8.598 & 665.63\\
\cline{2-7}        & 9 & 41008029 & 1.2976237e+09 & 1.3386317e+09 & 8.598 & 4390.17\\
\hline          & 2 & 725 & 1036 & 2481 & 8.501 & 0.04\\
\cline{2-7}  & 3 & 8305 & 21980 & 38562 & 8.501 & 0.12\\
\cline{2-7}          & 4 & 249439 & 1149753 & 1648373 & 8.501 & 3.75\\
\cline{2-7}        & 5 & 4339737 & 28293352 & 36972211 & 8.501 & 83.32\\
\cline{2-7} {\tt Liveness} & 6 & 12678951 & 1.1096373e+08 & 1.3851011e+08 & 8.501 & 314.08\\
\cline{2-7}        & 7 & 20128894 & 2.0273546e+08 & 2.4108713e+08 & 8.501 & 531.80\\
\cline{2-7}        & 8 & 25361336 & 3.4848047e+08 & 3.8927174e+08 & 8.598 &  936.05\\
\cline{2-7}        & 9 & 40305514 & 2.307274e+09 & 2.3482827e+09 & 8.598 & 7039.02\\
\hline \hline
\end{tabular*}
\end{center}
\end{table}

In Table~\ref{tab3} we describe the number of states and time required to prove the {\tt safety} and {\tt liveness} properties for the TTA Startup protocol using {\sf SAL-smc}. We have been able to verify both {\tt safety} and {\tt liveness} properties for TTA startup protocol for upto 8 nodes in around 1 hour. Let us contrast our verification effort
with the dense time modeling and verification of the same protocol reported in~\cite{DS04a, DS04b}. Using bounded model checking the same {\tt safety} property was proved for only $2$ nodes by $k$-induction at depth $8$, that too using $3$ auxiliary lemmas (the proof failed for $3$ nodes).
However, the invariant can be strengthened by constructing an abstraction of the transition systems using a verification diagram-based approach~\cite{Rus00}, and subsequently the property was verified for upto $10$ nodes.
\begin{table}
 \caption{Computational resources required to verify {\tt safety} and {\tt liveness} property by {\sf SAL-smc} for the TTA Startup}\label{tab3}
\begin{center}
{\footnotesize
\begin{tabular*}{0.5\textwidth}{@{\extracolsep{\fill}}||c|c|c|c||}
\hline
\hline
\# Nodes  & \# States & Time ({\tt Safety}  & Time ({\tt Liveness}\\
& & Property) (sec) &  Property) (sec) \\
\hline
2 & 68 & 0.34 & 1.18\\ \hline
3 & 485 & 0.63 & 3.28\\ \hline
4 & 5297 & 2.75 & 10.56\\ \hline
5 & 76345 & 13.11 & 48.31\\ \hline
6 & 1331650 & 77.23 & 563.82\\ \hline
7 & 26872795 & 4044.31 & 742.90\\ \hline
8 & 615902175 & 3440.63 & 3101.26\\
\hline
\hline
\end{tabular*}}
\end{center}
\end{table}

\section{Extension of Timeout and Calendar based Models} \label{extend}
In this section we extend our model to incorporate other modeling concepts like inter-process scheduling, priorities and interrupts, and urgent and committed locations. These extensions will be illustrated using ToM as a base model, however they can be easily adapted for calendar based ToM also. Also note that the digitization result presented in Section~\ref{digi_tom}, and the finitary reduction and associated clockless modeling proposed in Section~\ref{clockless} are applicable to these extended models as well because the additional components defined in these (extended) models are independent of the variables present in the base model and therefore, do not affect the underlying semantics of the base model.

\subsection{Modeling Inter-Process Scheduling}
So far, we have considered models capturing true parallelism with non-determinism. However, in some cases the ability of a system to meet real-time constraints crucially depends on the number of processors that are available and also, on the process scheduling algorithm. Thus, we need to distinguish between the models of multiprocessing and multiprogramming. We show how ToM can be extended to include fixed number of programs that are executed by time sharing, on a single processor. Subsequently we use our framework to model priorities and interrupts for a general distributed multiprogramming system. These are motivated by the framework of multiprogramming system introduced in~\cite{HMP92a}.

A Multiprogramming Timeout based Model (MToM) $P$ has the form
$$ \{ \theta\}[(P_{11} ||| \ldots ||| P_{1l_1})||(P_{21} ||| \ldots ||| P_{2l_2})||\ldots || (P_{m1} ||| \ldots ||| P_{ml_m})], $$
where each process $P_{i1}\ldots P_{il_i}, 1 \leq i \leq m$ is a sequential non-deterministic process as we have seen before. By $P_\alpha ||| P_\beta$ we mean processes $P_\alpha$ and $P_\beta$ share a single processor and are executed on one transition at a time according to some scheduling policy. Thus there are $m$ groups of processes in the above MToM such that all the processes in a group share the same processor, e.g., the processes $P_{11} \ldots P_{1l_1}$ would execute on the first processor. Processes in different groups running on different processors execute concurrently as in the case of ToM defined in Section~\ref{tom_syntax}. A special case of synchronous communication needs special care because both the processes need to be simultaneously active: If process $P_{ij}$ and $P_{i'j'}$ have a synchronous communication, these processes must be executing on different processors, that is, $i \neq i'$.

For example, $[(P_{11} ||| P_{12} ||| P_{13})|| (P_{21}||| P_{22})]$ is the model of a system with five processes running on two processors. The first three processes share the first processor and next two the second processor. A synchronous communication can take place between two processes only when these processes belong to different groups. 

A timed transition system $S_P = (\V, \Sigma, \Sigma_0, \Gamma)$ can be associated with an MToM also. The key difference now is that $\V$ contains additional processor control variables $\mu_1,\ldots, \mu_m$, such that $\mu_i$ ranges over $\{1, \ldots, l_i, \perp \}$, {\em i.e.}, $\V = \U \cup \{ \mu_1,\pi_{11}, \ldots, \pi_{1l_1}\} \cup \{\mu_2,\pi_{21}, \ldots, \pi_{2l_2}\} \cup \ldots \cup \{\mu_m, \pi_{m1},\ldots, \pi_{ml_m}\}$. The processor control variables assume the value $\perp$ before the processor starts executing the processes in a group. Thereafter, the control of the process $P_{i\mu_i}$ resides at the location ${\pi}_{\mu_i}$ executing on the $i^{th}$ processor. In other terms, only the process $P_{i\mu_i}$ is active on the $i^{th}$ processor, while all other processes $P_{ij}, j \ne \mu_i$ are suspended. When the execution of the process $P_{i\mu_i}$ is suspended as per the scheduling policy, in future it can only resume at the last suspended location $\pi_{i\mu_i}$.


For simplicity, we will next consider the case of a single processor, that is $m = 1$ and will drop the subscript $1$ in the notations e.g., $\mu$ would stand for $\mu_1$ and $\pi_{j}$ for $\pi_{1j}$. Let us now discuss some of the transitions that would additionally occur in this framework. For example, $\Gamma$ will contain a set of scheduling transitions, $\Gamma_{sch}$.


A scheduling policy determines the set of scheduling transitions. We consider only scheduling policies with a single entry transition, that is enabled on all states. The entry transition is assumed to be enabled on the initial states, and activates non-deterministically one of the competing processes. A very popular and simple scheduling policy is based on {\it greedy scheduling}. According to which, a process, currently in the control of the processor, continues to remain active until all its transition are disabled, when an arbitrary (other) process  with an enabled transition takes over. More flexible scheduling strategies can be implemented by incorporating explicit scheduling instruction $resume(s)$, where $s \subset \{1, \ldots,n\}$ determines a subset of processes. The scheduling operation $resume(s)$ suspends the currently active process, $P_i$ and activates, nondeterministically, one of the processes $P_j$, with $j \in s$. A scheduling edge in the process $P_i$ will be represented as:
$$l_j^i\ \stackrel{\rho\ \Rightarrow \lan resume(s), [l, m] \ran}{\longrightarrow}\ l_k^i$$
Where $[l, m], l < m$ specifies (optional) delay which the scheduling operation may take between $l$ and $m$ time units. Such an edge introduces an additional transition in $\Gamma$, and grouped in $\Gamma_{sch}$ as follows:
\[
\nu_{sch} \equiv (\sigma, \sigma') \in \Gamma_{sch} \Leftrightarrow \left\{
\begin{array}{ll}
1.\, & \rho \mbox{ holds in } \sigma \\
2.\, & \sigma'(t) = \sigma(t) + \delta \\
3.\, & \forall y \in \V \setminus \{\mu, \pi_i\}: \sigma'(y) = \sigma(y) \\
4.\, & \sigma(\mu) = i \mbox{ and } \sigma'(\mu) \in s\\
5.\, & \sigma(\pi_i) =\ l_j^i \mbox{ and } \sigma'(\pi_i) = l_k^i
\end{array} \right.
\]
Where $\delta$ is a randomly selected constant such that $l \leq \delta \leq m$. To add, a $suspend(i, j)$ operation, which suspends a process $P_i$ and activates process $P_j$, can also be defined as $resume(\{1 \leq j \leq m \mid i \neq j\})$, that is, the instruction $suspend(i, j)$ delegates the control from the currently active process $P_i$ to the process $P_j$. In practice, processes $P_i$ and $P_j$ could have some operational relationship with each other, e.g., $P_i$ is the parent process, which spawns $P_j$ as its its child process, goes into waiting state and activates $P_j$. On termination $P_j$ may hand over the control back to $P_i$ using the operation $resume(\{i\})$.

\subsection{Modeling Priorities and Interrupts}

We will next discuss how interrupts can be handled by way of introducing static priorities with global preemption semantics. Priorities will be represented using non negative integers and will be assigned to every transition such that lower value would be interpreted as higher priority. During execution a transition with the highest priority at any time point is selected and current process would be suspended if the ready process having the transition with the highest priority happens not to be the current process. A Multiprogramming Timeout based Model (MToM) $P$ with priority is one in which a priority is associated with every transition in the timed transition systems for $P$. Using priorities it is possible to design a simple, static scheduling strategy without resorting to explicitly constructing a scheduler.

As an example, in a ToM, an extended timeout edge $e: (l_j^i, \rho_e\  \Rightarrow \lan \tau_i := update_i, \gamma, f, \mathbf{p_{e}} \ran, l_k^i)$ in the graph of the process $P_i$  would be represented as $$e: l_j^i\ \stackrel{\rho_e\ \Rightarrow \lan \tau_i := update_i, \gamma, f, \mathbf{p_{e}} \ran}{\longrightarrow}\ l_k^i,$$ where an additional parameter $\mathbf{p_{e}} \in \N$ is the priority associated with the transition $e$. All other edges e.g., synchronous communication and asynchronous communication would be extended similarly.

Accordingly, we extend the semantics also. For the prioritized timeout edges, a transition with the highest priority is allowed by adding it in $\Gamma_0$ in the following way.\\ \\
{\sf Prioritized Timeout Increment Transition:} Collect all those extended timeout edges $e$ for which corresponding transitions are enabled in the current state $\sigma$, that is, $\rho_e$ \textit{holds in} $\sigma$. Let $En_{\sigma}$ be the set of these enabled edges. Now select those timeout edges $e_h \in En_{\sigma}$, which have the highest priority, i.e., $\forall {e'} \in En_{\sigma}. \mathbf{p_{h}} \leq \mathbf{p_{e'}}$. Add transition $\nu_h \equiv (\sigma, \sigma')$ in $\Gamma_0$  such that:
\[
\nu_h \equiv (\sigma, \sigma') \in \Gamma_0 \Leftrightarrow \left\{
\begin{array}{ll} 1.\, &
\rho_h \mbox{ holds in } \sigma \\
2.\, & \sigma'(t) = \sigma(t) \\
3.\, &  \mathbf{If}\, \sigma(\tau_i) = \sigma(t)\\
& \ \ \ \mathbf{then }\, \sigma'(\tau_i) = update_i > \sigma(\tau_i)\\
& \mathbf{else}\, \sigma'(\tau_i) = \sigma(\tau_i)\\
4.\, & \forall y \in \gamma:\ \sigma'(y) = \sigma_e(t)\ \mbox{ and}\\
& \forall x \in \X\setminus\gamma:\ \sigma'(x) = \sigma(x) \\
5.\, & \forall v \in G \cup L_i: \sigma'(v) = f(\sigma(v))\ \mbox {and}\\
& \forall v \in \mathit{Var}\setminus(G \cup L_i):\ \sigma'(v) = \sigma(v)\\
6.\, & \sigma(\pi_i) =\ l_j^i \mbox{ and } \sigma'(\pi_i) = l_k^i \\
7.\, & \sigma'(\mu) = i
\end{array}
\right.
\]
If there are multiple enabled edges with the same highest priority, their corresponding transitions are non deterministically interleaved.

The remaining all other transitions can also be extended similarly. Under such extended syntax and semantics, an interrupt can be modeled as an edge having relatively high priority than other enabled transitions: $$e_{int}: (l_j^i, True\  \Rightarrow \lan \tau_i := update_i, f, \mathbf{p_{int}} \ran, l_k^i)$$ where $update_i$ specifies the delay in interrupt processing and $f$ specifies the steps in interrupt processing. Note  $\mathbf{p_{int}}$ is such that $\forall \sigma \in \Sigma . \forall e \in En_{\sigma}. \mathbf{p_{int}} \leq \mathbf{p_{e}}$.

\subsection{Modeling Urgent Location and Committed Location}

In UPPAAL there are three different types of locations: \textit{normal locations}, \textit{urgent locations} and \textit{committed locations}~\cite{BDL04}. In a normal location time can progress, but in urgent and committed locations time is not allowed to proceed. Moreover, there is a subtle difference between urgent and committed locations. Urgent locations can be interleaved with the normal locations, but a committed location has to be followed by its immediate successor. The requirement of considering a location to be urgent or committed arises out of the nature of the application being modeled in UPPAAL. For example, committed locations are used to model atomic behaviors in multi-way synchronizations and atomic broadcasting in real-time systems~\cite{BGK+02}.

In timeout and calendar based models, we model an urgent or a committed location in the following way. For all the incoming edges to the the urgent or committed location in process $P_i$, $update_i$ is set to current time $t$, and in case of clockless modeling $update_i$ is set to $0$.

If a process in a system has a committed location, we introduce a boolean variable $committed\_flag$ in the set of global variables $G$. For all the incoming edges to a committed location, $committed\_flag$ is set to $1$ (part of $f$) and for incoming edges to a non-committed state one is not allowed to set the flag to $1$. The guard $\rho$ for a transition following a committed location is always \textit{True} and $committed\_flag$ is reset during this transition. For all the transitions except those following the committed locations, the existing guard $\rho$ is replaced by $\rho \wedge (committed\_flag \neq 1)$. This will not allow any other process to take a discrete transition when a process is in a committed state.

\section{Conclusion and Further Work}\label{conclude}

In this work we have considered the well-known problem of real-time verification with dense time dynamics using timeout and calendar based models and proposed a technique to simplify this to a finite state verification problem. Towards this, we define a specification formalism for these models as timeout transition diagrams with associated transition system semantics. Next, we proposed a two-step reduction technique for rendering these models amenable to finite state verification under discrete dynamics. Our experimental results bring out the advantages gained by this technique over infinite state modeling and verification. Experiments on Fisher's protocol and TTA startup protocol highlight that the verification technique scales reasonably well. Further, {\sf liveness} properties can be verified in this framework, which is beyond the capability of infinite state verification. Though in~\cite{DS04a}, it has been reported that verification of Fischer's protocol can be scaled up to $53$ nodes, the verification process involved finding out auxiliary lemmas  manually, which is a non-trivial process. On the other hand our finite state verification, though could not be scaled to this extent, is nonetheless simple and straight-forward. The verification effort involves only modeling the protocols faithfully. {\sf SAL} offers a number of tools for finite state verification, for example, {\sf SAL-sim}, {\sf SAL-path-finder} and {\sf SAL-deadlock-checker}, which help quite a lot in the verification process. Such tool support is yet not available for infinite state verification. Moreover, one can use any finite state verification engine of choice using our framework.

We limited our attention to the qualitative temporal properties that exclusively corresponds to LTL formulas. However, the proposed reduction technique is amenable to any specification logic which is closed under inverse digitization including branching time temporal logics CTL or CTL$^\ast$.


The effectiveness of the proposed finitary reduction technique can be further scaled up by integrating it with additional abstraction techniques to verify parametric systems,  with arbitrary but finite number of identical processes. Sa\"{\i}di and Lesens~\cite{LS97} presented an algorithm for automatically constructing abstraction for such systems to verify {\tt safety} properties. The $(0,1, \infty)$ counter abstraction method proposed in~\cite{PXZ02} deals with the verification of {\tt liveness} properties by abstracting a parameterized system of unbounded size into a finite-state system. The proposed formalism can be further optimized by considering timeouts as shared variables among processes, so that timeout updation rules could specify new timeout values based upon those of other processes in the system. This optimization would increase the level of synchronization between component processes and would hopefully scale up the models.


In the larger perspective it can be said that for most of the timeout and calendar based models (i.e., for which timeout updates are not restricted to $(0,1)$-interval) verification of LTL properties with dense time dynamics reduces to finite state modeling and verification of the same properties. In industrial designs, this could offer a significant advantage as it is easier for practitioners to use finite state model checkers to model and verify timed systems.

Decidability and complexity theoretic aspects of the reachability analysis on these models is an important research direction for further investigation. A comparison of expressiveness of  ToM (or calender based ToM) with other known formal models of real-time systems including Timed Automata~\cite{Alur99}, Timed Petri Nets~\cite{JW98}, and Timed Process Algebras~\cite{BB91} would shed light on the comparative strength of these models for practical purposes. For example, these comparisons could reveal other properties desirable of a modeling framework including compositionality, robustness against clock drifts, and may demonstrate the difficulty of modeling timeout models using these models as compared to ToM.

\bigskip
\noindent
{\bf Acknowledgment} Indranil Saha and Suman Roy did most of this work when they were with HTS Research, Bangalore.


\begin{thebibliography}{Lam94}

\bibitem[Alu99]{Alur99} R. Alur. ``Timed Automata''. In {\em Proceedings of International Conference on Computer-Aided Verification (CAV'99)}, LNCS 1633, Springer-Verlag, pp. 8-22, 1999.

\bibitem[AlD94]{AD94} R. Alur and D. L. Dill. ``A Theory of Timed Automata.'' In {\em Theoretical Computer Science}, vol. 126, number 2, pp. 183-–235, 1994.


\bibitem[AlH91]{AH91} R. Alur and T.A. Henzinger. ``Logics and Models of Real Time: A Survey.'' In {\em Proceedings of the Real-Time: Theory in Practice, REX Workshop}, pp. 74-106, Springer, 1991.

\bibitem[BMN00]{BMN00} P. Bellini, R. Mattolini, and P. Nesi. ``Temporal Logics for Real-Time System Specification.'' In {\em ACM Computing Surveys}, vol. 32, number 1, 2000.

\bibitem[BeJ91]{BB91} Jos C. M. Baeten and Jan A. Bergstra. ``Real Time Process Algebra.'' In {\em Formal Aspects of Computing}, vol. 3, number 2, pp. 142--188, 1991.

\bibitem[BDL04]{BDL04} Behrmann G., David A. and Larsen K. G.. ``A Tutorial on UPPAAL''. In \emph{4th International School on Formal Methods for the Design of Computer, Communication, and Software Systems (SFM-RT'04)}, LNCS 3185, Springer-Verlag, pp. 200--236, 2004.

\bibitem[BGK02]{BGK+02} J. Bengtsson, W. O. D. Griffioen, K. J. Kristoffersen, K. G. Larsen, F. Larsson, P. Pettersson and W. Yi. ``Automated Analysis of an Audio Control Protocol Using Uppaal''. In {\em Journal of Logic and Algebraic Programming},  vol. 52-53,  Holger Hermanns and Joost-Pieter Katoen (eds.), pp. 163--181, July-August, 2002.

\bibitem[BLN03]{BLN03} Beyer, D., Lewerentz, C., and Noack, A. Rabbit: A tool for BDD-based verification of real-time systems. In {\em Computer Aided Verification}, pp. 122--125, Springer, 2003.

\bibitem[Bozga]{BDM+98} Bozga, M., Daws, C., Maler, O., Olivero, A., Tripakis, S., Yovine, S. ``Kronos: A model-checking tool for real-time systems''. In {\em Computer Aided Verification}, pp. 546--550, Springer, 1998.

\bibitem[Bos99]{Bos99} D. Bo$\breve{s}$na$\breve{c}$ki. ``Digitization of Timed Automata''. In {\em Proceedings of the Fourth International Workshop on Formal Methods for Industrial Critical
Systems (FMICS'99)}, Trento, Italy, pp. 283--302, 1999.

\bibitem[BoD98a]{BD98a} D. Bo$\breve{s}$na$\breve{c}$ki and D. Dams. ``Integrating Real Time into Spin: A Prototype Implementation''. In {\em Proceedings of the Formal Description Techniques and Protocol Specification, Testing and Verification (FORTE/PSTV'98)}, Kluwer, pp. 423--439, 1998.

\bibitem[BoD98b]{BD98b} D. Bo$\breve{s}$na$\breve{c}$ki and D. Dams. ``Discrete-Time PROMELA and Spin''. In {\em Proceedings of Formal Techniques in Real-Time and Fault-Tolerant Systems (FTRTFT'98)}, LNCS 1486, Springer, pp. 307--310, 1998.

\bibitem[ChH04]{CH04} Chun K.Y., Hung D.V. ``Verifying Real-Time Systems using Untimed Model Checking Tools''. In {\em Technical report UNU-IISTTR-3002}, The United Nations University, International Institute for Software Engineering, 2004.

\bibitem[CGP99]{GGP99} E. M. Clarke, O. Grumberg, and D. A. Peled. ``Model Checking''. The MIT press, 1999.

\bibitem[CHR91]{dc91} Z. Chaochen, C.A.R. Hoare, and A. P. Ravn. ``A Calculus of Duration.'' In {\em Information Processing Letters}, vol. 40, pp. 269-–276, 1991.

\bibitem[DaS95]{DS95} J. Davies, and S. Schneider. ``A Brief History of Timed CSP''. In {\em Theoretical Computer Science}, vol. 138, number 2, pp. 243--271, 1995.

\bibitem[DuS04a]{DS04a} B. Dutertre and M. Sorea. ``Timed Systems in SAL''. Technical Report, Computer Science Laboratory, SRI International, 2004.

\bibitem[DuS04b]{DS04b} B. Dutertre and M. Sorea. ``Modeling and Verification of a Fault-Tolerant Real-Time Startup Protocol using Calendar Automata''. In {\em Formal Techniques in Real-Time and Fault-Tolerant Systems (FTRTFT'04)}, LNCS 3253, Springer-Verlag, pp. 199--214, 2004.

\bibitem[HMP92a]{HMP92} T. A. Henzinger, Z. Manna, and A. Pnueli. ``What Good Are Digital Clocks?''. In \emph{Proceedings of the 19th International Colloquium on Automata, Languages,
and Programming (ICALP'92)}, LNCS 623, Springer-Verlag, pp. 545--558, 1992.

\bibitem[HMP92b]{HMP92a} T. A. Henzinger, Z. Manna, and A. Pnueli. ``Timed transition Systems''. In \emph{Proceedings of the Real-Time: theory in Practice, REX Workshop} . J. W. Bakker, C. Huizing, W. P. Roever, and G. Rozenberg, (Eds.) LNCS 600, Springer-Verlag, pp. 226--251, 1992.

\bibitem[Hol93]{Hol03} G. J. Holzmann. ``The Spin Model Checker, Primer and Reference Manual''. Addison-Wesley, 2003.

\bibitem[LeS97]{LS97} D. Lesens and H. Sadi. ``Automatic Verification of Parameterized Networks of Processes by Abstraction.'' In \emph{International Workshop on Verification of Infinite State Systems (INFINITY'97)}, 1997.

\bibitem[MRS03]{MRS03} L. M. Moura, H. Ruess, and M. Sorea. ``Bounded Model Checking and Induction: From Refutation to Verification''. In Proceedings of {\em Computer-Aided Verification}, LNCS 2725, Springer-Verlag, pp. 14--26, 2003.

\bibitem[MOR04]{MOR+04} L. M. Moura, S. Owre, H. Ruess, J. M. Rushby, N. Shankar, M. Sorea, and A. Tiwari. ``Sal 2''. In {\em Proceedings of International Conference on Computer-Aided Verification (CAV'04)}, LNCS 3114, Springer, pp. 496--500, 2004.

\bibitem[NiS94]{NS94} X. Nicollin, and J. Sifakis. ``The Algebra of Timed Processes, ATP: Theory and Application.'' In {\em Information and Computation}, vol. 114, number 1, pp. 131--178, 1994.

\bibitem[Pik05]{pike05} L. Pike. ``Real-Time System Verification by $k$-Induction''. Technical report, NASA Langley Research Center. TM-2005-213751, 2005. Available at {\sf http://www.cs.indiana.edu/~lepike/pub\_pages/reint.html}

\bibitem[Pnu77]{Pn77} A. Pnueli. ``The Temporal Logic of Programs''. In {\em Proceedings of the 18th Annual Symposium of Foundations of Computer Science}, IEEE Computer Society Press, pp. 46--57, 1977.

\bibitem[OsN96]{ON96} J. Ostro and H. Ng. ``Verifying Real-Time Systems with Standard Tools.'' In {\em AMAST workshop on real-time systems}, 1996. Available at {\sf http://www.cse.yorku.ca/~stateclock/StateClock/amast.pdf}

\bibitem[PXZ02]{PXZ02} A. Pnueli, J. Xu, and L. Zuck. ``Liveness with (0,1,infinity)-Counter Abstraction.'' In \emph{Proceedings of International Conference on Computer-Aided
Verification (CAV'02)}, LNCS 2404, Springer-Verlag, pp. 107--122, 2002.

\bibitem[Rus00]{Rus00} J. Rushby. ``Verification Diagrams Revisited: Disjunctive Invariants for Easy Verification''. In {\em Proceedings of International Conference on Computer-Aided Verification (CAV'00)}, LNCS 1855, Springer-Verlag, pp. 508--520, 2000.

\bibitem[SaR06a]{SR06a} I. Saha and S. Roy. ``A Finite State Modeling of AFDX Frame Management using Spin''. In {\em Proceedings of the 11th International Workshop on Formal Methods for Industrial Critical Systems (FMICS'06)}, 2006.

\bibitem[SaR06b]{SR06b} I. Saha and S. Roy. ``A Finite State Analysis of Time-triggered CAN (TTCAN) Protocol using Spin''. In {\em Proceedings of the International Conference on Computing: Theory and Application (ICCTA'07)}, IEEE Computer Society, 2007.

\bibitem[SMR07]{SMR07} I. Saha, J. Misra, S. Roy. ``Timeout and Calendar based Finite State Modeling and Verification of Real-Time Systems''. In \emph{Proceedings of Automated Technology for Verification and Analysis (ATVA'07)}, LNCS 4762, pp. 284--299, 2007.

\bibitem[Ste05]{Ste05} Steiner, W. ``Model-Checking Studies of the FlexRay Startup Algorithm''. Technische Universit{\"a}t Wien, Institut f{\"u}r Technische Informatik, Treitlstr. 1-3/182-1, 1040 Vienna, Austria, Research Report, 57/2005, 2005.

\bibitem[StP02]{SP02} W. Steiner and M. Paulitsch. ``The Transition from Asynchronous to Synchronous System Operation: An Approach for Distributed Fault-Tolerant System''. In {\em Proceedings of the 22nd International Conference on Distributed Computing Systems (ICDCS'02)}, IEEE Computer Society, pp. 329--336, 2002.

\bibitem[SRS04]{SRSP04} Steiner, W. and Rushby, J. and Sorea, M. and Pfeifer, H. ``Model Checking a Fault-Tolerant Startup Algorithm: From Design Exploration To Exhaustive Fault Simulation''. In {\em Proceedings of DSN}, 2004.

\bibitem[TrC96]{TC96} S. Tripakis and C. Courcoubetis. ``Extending PROMELA and Spin for Real Time''. In {\em Proceedings of the Second International Workshop on Tools and Algorithms for the Construction and Analysis of Systems, (TACAS'96)}, LNCS 1055, Springer Verlag, pp. 329--348, 1996.

\bibitem[Jia98]{JW98} Jiacun Wang. ``Timed Petri Nets: Theory and Application.'' Kluwer Acadamic Publishers, USA, October 1998.

\bibitem[VaW86]{VW86} M. Y. Vardi and P. Wolper. ``An Automata-Theoretic Approach to Automatic Program Verification''. In {\em Proceedings of 1st IEEE Symposium on Logic in Computer Science (LICS'86)}, IEEE Computer Society Press, pp. 332-344, 1986.

\end{thebibliography}
\end{document}